\documentclass[aps,pra,superscriptaddress,twocolumn,longbibliography]{revtex4-1}
\usepackage{amsmath}
\usepackage{bbm}
\usepackage[normalem]{ulem}
\usepackage{times}
\usepackage[normalem]{ulem}
\usepackage{graphicx}
\usepackage{dcolumn}
\usepackage{bm}
\usepackage{tabularx}
\usepackage{enumerate}

\usepackage{amsthm}
\usepackage{amsfonts,amssymb}
\usepackage{mathrsfs}
\usepackage{subfigure}
\usepackage{pifont}
\usepackage{color}
\usepackage{thmtools}
\usepackage{verbatim}
\usepackage[ruled,vlined]{algorithm2e}
\usepackage{qcircuit}

\newtheorem{theorem}{Theorem}
\newtheorem{lemma}{Lemma}
\newtheorem{definition}{Definition}
\newtheorem{proposition}{Proposition}
\newtheorem{observation}{Observation}

\newcommand{\bra}[1]{\langle#1|}
\newcommand{\braket}[1]{\langle#1\rangle}

\newcommand{\ket}[1]{|#1\rangle}

\newcommand{\expectv}[1]{\langle#1\rangle}
\newcommand{\dcut}{\Delta_{\mathrm{cut}}}

\newcommand{\spacespan}{\text{span}^{\mathbf{R}} }

\newcommand{\var}{\mathrm{Var}}

\newcommand{\mc}[1]{\mathcal{#1}}

\newcommand{\ketbra}[2]{\langle#1|#2\rangle}

\usepackage{hyperref}
\hypersetup{colorlinks=true, linkcolor=black, citecolor=blue, urlcolor=blue}

\newcommand{\sustech}{Department of Physics, Southern University of Science and Technology, Shenzhen 518055, China}
\newcommand{\siqse}{Shenzhen Institute for Quantum Science and Engineering, Southern University of Science and Technology, Shenzhen 518055, China}
\newcommand{\pku}{Center on Frontiers of Computing Studies, Peking University, Beijing 100871, China}

\begin{document}

\title[]{Low-depth Hamiltonian Simulation by Adaptive Product Formula}

\author{Zi-Jian Zhang} 
\email{zi-jian@outlook.com}
\thanks{These two authors contributed equally }

\affiliation{\sustech}
\affiliation{\siqse}
\affiliation{\pku}

\author{Jinzhao Sun}
\email{jinzhao.sun@physics.ox.ac.uk}
\thanks{These two authors contributed equally }
\affiliation{Clarendon Laboratory, University of Oxford, Parks Road, Oxford OX1 3PU, United Kingdom}

\author{Xiao Yuan}
\email{xiaoyuan@pku.edu.cn}
\affiliation{\pku}

\author{Man-Hong Yung}
\email{yung@sustech.edu.cn}
\affiliation{\sustech}
\affiliation{\siqse}
\affiliation{Shenzhen Key Laboratory of Quantum Science and Engineering, Southern University of Science and Technology, Shenzhen 518055, China}

\date{\today}

\begin{abstract}
Various Hamiltonian simulation algorithms have been proposed to efficiently study the dynamics of quantum systems on a quantum computer. The existing algorithms generally approximate the time evolution operators, which may need a deep quantum circuit that is beyond the capability of near-term noisy quantum devices. 
Here, focusing on the time evolution of a fixed input quantum state, we propose an adaptive approach to construct a low-depth time evolution circuit. By introducing a measurable quantifier that characterizes the simulation error, we use an adaptive strategy to learn the shallow quantum circuit that minimizes that error. 
We numerically test the adaptive method with electronic Hamiltonians of the $\mathrm{H_2O}$ and $\mathrm{H_4}$ molecules, and the transverse field Ising model with random coefficients. 
Compared to the first-order Suzuki-Trotter product formula, our method can significantly reduce the circuit depth (specifically the number of two-qubit gates) by around two orders while maintaining the simulation accuracy.
We show applications of the method in simulating many-body dynamics and solving energy spectra with the quantum Krylov algorithm. 
Our work sheds light on practical Hamiltonian simulation with noisy-intermediate-scale-quantum devices.
\end{abstract}

\maketitle

\emph{\textbf{Introduction}.---}
One major application of quantum computing is to simulate the time evolution of many-body quantum systems~\cite{feynman1982simulating}, which not only allows us to study dynamical behaviors of quantum systems~\cite{georgescu2014quantum,buluta2009quantum,blatt2012quantum,mcardle2018quantum,lewis2019dynamics,motta2020determining,sun2021quantum,sun2021perturbative,yuan2020quantum,altman2019quantum}, but also constitutes other quantum algorithms as a common subroutine (e.g.~quantum phase estimation~\cite{kitaev1995quantum,Aspuru-Guzik1704}) for various tasks~(e.g.~finding energy spectra~\cite{stair2020multireference,motta2020determining, parrish2019quantum,somma2019quantum,kyriienko2020quantum,endo2020variational,2021arXiv210915304Z}). Various Hamiltonian simulation algorithms have been proposed, from the initial Trotter-Suzuki product formula~\cite{suzuki1985decomposition,suzuki1991general} to the latest advanced approaches~\cite{berry2015hamiltonian,childs2018toward,kivlichan2018quantum,berry2020time,childs2019theory}, such as ones based on Taylor series~\cite{PhysRevLett.114.090502} and quantum signal processing with qubitization~\cite{low2019hamiltonian,low2017optimal}. 
While these advanced approaches have theoretically improved scaling in the asymptotic limit,  product formula methods perform surprisingly well than what is expected from the worst-case bound.
Theoretically, tighter bounds have been derived for special types of systems, such as lattice Hamiltonians with nearest-neighbor interactions~\cite{childs2019nearly,childs2019theory}, verifying the empirical error estimates of the product formula methods. Numerically, it has been shown that product formula methods require orders of fewer entangling gates and T gates than the advanced methods~\cite{childs2018toward,childs2019theory}. Since product formula methods also require fewer qubits and are much easier to implement in experiments, they are favored for simulating the dynamics of intermediate-scale quantum systems. 

Consider a Hamiltonian $H=\sum_{j=1}^{L} a_j P_j$ decomposed as a sum of tensor products of Pauli operators $P_j$ with real coefficients $a_j$. The key idea of the product formula (PF) methods is to approximate the time evolution operator $e^{-iH\delta t}$ via a sequence of operators selected from $\{e^{-iP_j \delta t'}\}$, such as the first-order Suzuki-Trotter formula $e^{-iH\delta t} \approx \prod_{j=1}^{L}e^{-ia_jP_j \delta t}+\mc{O}(\delta t^2)$. More advanced approximation methods include higher-order Suzuki-Trotter formula~\cite{suzuki1991general}, randomization of the operator order~\cite{childs2019faster}, and qDRIFT via importance sampling~\cite{PhysRevLett.123.070503}. Since product formula methods approximate the time evolution operator (quantum channel), which works for arbitrary input states, they may require an unnecessarily large number of gates when we only evolve a specific quantum state in practice. Recent numerical and theoretical studies showed that product formula with fixed or random input states only needs a much shallower circuit~\cite{heyl2019quantum,chen2020quantum,zhao2021hamiltonian}. On the one hand, this indicates even stronger practicality of the product formula methods for fixed input states. On the other hand, since existing product formula methods do not exploit the input state  information, whether we can further exploit such information to reduce the implementation complexity deserves further study. As near-term quantum hardware has limited gate fidelity \cite{preskill2018quantum,gu2017microwave,kjaergaard2019superconducting,arute2019quantum,krantz2019quantum,kandala2017hardware,kandala2019error,bharti2021noisy}, it is crucial to design Hamiltonian simulation algorithms with lower circuit depth and hence higher calculation accuracy. 

In this work, we propose a learning/construction-based method to adaptively find an optimized product formula for evolving an unknown but fixed input quantum state. Instead of using the theoretical worst-case error bound of conventional product formula methods, we introduce a measurable quantifier to describe the simulation error. We consider different evolution operators at different time and construct the optimal one by minimizing the error quantifier. Since the error quantifier focuses on the quantum state instead of the entire evolution channel, we can thus adaptively obtain the state evolution circuit with a significantly reduced circuit depth than conventional approaches for the unitary evolution.
We numerically verify our method by considering the time evolution of molecular systems and spin systems and show reductions of CNOT gate count by around two orders compared to the first-order Trotter formula. \\

\emph{\textbf{Adaptive product formula (single step)}.---}
Now we briefly review the conventional product formula and introduce our adaptive approach that exploits the input state information. We consider time-independent Hamiltonian 
$H=\sum_{j=1}^{L} a_j P_j$ with $L$ terms of Pauli words $P_j$, which are tensor products of Pauli operators. The product formula approaches approximate the time evolution operator $e^{-iH\delta t}$ within a small time step $\delta t$ as
\begin{equation}\label{Eq:firstpd}
    \varepsilon = \|e^{-iH\delta t  } -  \prod_j e^{-iO_j\lambda_j\delta t}\|, 
\end{equation}
where $\varepsilon$ is the approximation error,
$O_j\in\{P_j\}$ are chosen from Pauli words in the Hamiltonian, $\lambda_j$s are real coefficients, and $\|\cdot\|$ is the operator norm. By properly choosing $\{O_j\}$ with a given order, the approximation error $\varepsilon$ can be suppressed to higher orders of $\delta t$ (e.g.~$\varepsilon =\mathcal{O}(\delta t^2)$) and we can accordingly simulate the whole evolution operator $e^{-iHT}$ with small error (e.g.~$\mathcal{O}(\delta tT)$). The error bound is a pessimistic estimation of the worst case scenario, and several numerical and theoretical studies~\cite{heyl2019quantum,chen2020quantum, zhao2021hamiltonian} have further shown that it could be much smaller given specific or random input quantum states.

Considering the state $\ket{\Psi(t)}$ at time $t$, 
instead of approximating the unitary operation $e^{-iH\delta t}$, we approximate the time-evolved state $e^{-iH \delta t}\ket{\Psi(t)}$ as 
\begin{equation}\label{eqn:eq1}
e^{-iH\delta t}\ket{\Psi(t)}\approx\prod_j e^{-iO_j\lambda_j \delta t}\ket{\Psi(t)},
\end{equation}
where $O_j$ are Pauli words of the Hamiltonian with real coefficients $\lambda_j$. 
We characterize the approximation with the Euclidean distance between the two states as
\begin{equation}\label{Eq:adaptiveep}
\begin{aligned}
     \tilde\varepsilon &= \| e^{-iH \delta t}\ket{\Psi(t)}-\prod_j e^{-iO_j\lambda_j \delta t}\ket{\Psi(t)} \|,
\end{aligned}     
\end{equation}
which,  when expanded to the first-order of $\delta t$, becomes $\tilde\varepsilon=\Delta \delta t + \mathcal{O}(\delta t^{3/2})$
with the first-order error 
\begin{equation}\label{eqn:distance}
\begin{aligned}
\Delta^2 =&\braket{ H^2 }+\sum_{j j'}  A_{jj'} {\lambda}_j {\lambda}_{j'} -2\sum_{j } C_j {\lambda}_j,     
\end{aligned}
\end{equation}
where the matrix elements $A_{jj'}=\text{Re}\left(\braket{\Psi(t)|O_jO_{j'}|\Psi(t)} \right)$, $C_j=\text{Re}\left( \braket{\Psi(t)|HO_j|\Psi(t)} \right)$, and $\braket{ H^2 }=\braket{\Psi(t)|H^2|\Psi(t)}$ could be efficiently measured via quantum circuits~\cite{NoteX}. 
Since the simulation error $\tilde\varepsilon$ is proportional to $\Delta$, 
it can be naturally regarded as a quantifier for the approximation of \autoref{eqn:eq1}.
Therefore, what we need is an approach to find the optimal $\{O_j\}$ and $\vec \lambda$ that minimize $\Delta$.
For this, we consider an adaptive strategy (\autoref{alg:one_step}).
Specifically,
we first try a series of Pauli words from the Hamiltonian. Then we calculate the optimal coefficients $\lambda_j$ that minimize  $\Delta^2$ by solving the linear equation
$
\sum_{j'} A_{jj'} \lambda_{j'} =  C_j
$~\cite{NoteX}. We only add the optimal Pauli word that gives the lowest $\Delta$ to the circuit.
We sequentially add more Pauli words until the error is lower than a given threshold $\dcut$.  
We summarize the method in \autoref{alg:one_step} and prove several features of this circuit construction strategy in~\autoref{thm:n_iter_bound}.
\begin{theorem}
\autoref{alg:one_step} satisfies the following properties. \\
(1) The error $\Delta$ strictly decreases at each iteration;\\
(2) Each Pauli word is only needed to appear once;\\
(3) We can achieve an error $\Delta\leq\dcut$ in at most $L$ iteration for any $\dcut \geq 0$.
\label{thm:n_iter_bound}
\end{theorem}

\begin{algorithm}[t]
\caption{Adaptive product formula (single step)}
\label{alg:one_step}
1. Set $\dcut$ and $n=1$. Input the initial state $\ket{\Psi(t)}$ and Hamiltonian $H=\sum_{j=1}^{L} a_j P_j$;

2. In the $n^{\mathrm{th}}$ iteration, calculate $A,C,\vec\lambda$ and $\Delta$ of the new circuit $e^{-iP_k\lambda_{k}\delta t}\prod_{j=1}^{n-1}e^{-iO_j\lambda_{j}\delta t}\ket{\Psi(t)}$ for each Pauli word $P_k$ in the Hamiltonian; 

3. Set $P_k$ that gives the lowest $\Delta$ in step 2 to be $O_n$;

4. If $\Delta > \dcut$ for the new circuit, go to step 2 with $n=n+1$. Else, stop and output $\vec O$, $\vec\lambda$ for approximation in \autoref{eqn:eq1}.
\end{algorithm}

\noindent The first property indicates that the strategy is always effective; The second property indicates that the strategy only requires applying each Pauli word  once~\footnote{This property only holds when we consider the first-order error $\Delta$, which may fail for higher-order errors.}; The third property states that \autoref{alg:one_step} will terminate in a finite number of steps. The above theorem thus guarantees the effectiveness and efficiency of the adaptive approach in the approximation of the time-evolved state in \autoref{eqn:eq1}. We leave the proof of \autoref{thm:n_iter_bound} to Supplemental Material.\\

\emph{\textbf{Adaptive product formula (jointly optimized)}.---} 
To evolve an initial state $\ket{\Psi_0}$ to $e^{-iHT}\ket{\Psi_0}=\ket{\Psi(T)}$, we can sequentially apply  the single-step adaptive strategy, in a similar vein to conventional product formula methods. However, this might not be optimal since the circuit constructed independently at each step may still be redundant for the state evolution.
Here we introduce a jointly-optimized adaptive strategy that produces a more compact quantum circuit.

Using an iterative description, we assume the approximation of the time-evolved state at time $t$ to be $\ket{\Psi(t)}=G(\vec O,\vec\Lambda)\ket{\Psi_0}$, where $G(\vec O,\vec\Lambda) =\prod_{j}e^{-iO_j\Lambda_j}$ is the circuit that has been constructed at time $t$. Here, we represent the list of Pauli words by $\vec O=\{O_j \} $ and the adjustable real coefficients (parameters) by $\vec \Lambda = \{\Lambda_j \} $.
The single-step adaptive strategy considers the approximation of \autoref{eqn:eq1} as $G(\vec O',\vec \lambda'\delta t)G(\vec O,\vec\Lambda) \ket{\Psi_0}$, in which the original circuit block $G(\vec O,\vec\Lambda)$ is fixed and only the new added circuit block $G(\vec O',\vec \lambda'\delta t)$ (specifically $\vec O'$ and the associated $\vec \lambda' $) is optimized.
Here, we propose a refined strategy (\autoref{alg:joint}) that jointly optimizes $G(\vec O',\vec \lambda'\delta t)$ and $G(\vec O,\vec\Lambda)$. That is, we consider the  approximation
\begin{equation}
    e^{-iH\delta t}\ket{\Psi(t)} \approx G(\vec O',\vec \lambda'\delta t)G(\vec O,\vec\Lambda+\vec \lambda\delta t) \ket{\Psi_0},
    \label{eq:jointapproxi}
\end{equation}
where we also optimize  $\vec\lambda$, which  correspond to the variation of the parameters found in the previous steps. 
Specifically, we consider the simulation error
$
\tilde\varepsilon = \|e^{-iH\delta t}\ket{\Psi(t)} - G(\vec O',\vec \lambda'\delta t)G(\vec O,\vec\Lambda+\vec \lambda\delta t) \ket{\Psi_0}\|$,
and the first-order error $\Delta=\lim_{\delta t\rightarrow 0} \tilde\varepsilon / \delta t$ has a similar form of Eq.~\ref{eqn:distance}. 
Given $\vec O', \vec O$ and $\vec \Lambda$, the optimal $\vec\lambda$ and $\vec\lambda'$ that minimize $\Delta$ could then be analytically obtained by solving a linear equation, and hence $\Delta$ can also be obtained (see \cite{NoteX} for the analytical  formula of $\Delta$ and detailed calculation of the optimal $\vec\lambda$ and $\vec\lambda'$).

Similar to the single-step strategy, the jointly-optimized strategy also constructs its circuit by iterating over all the Pauli words in the Hamiltonian. We calculate $\Delta$ for each Pauli word  and add the Pauli word that gives the lowest $\Delta$,  $O^{(\min)}$, to the previous circuit. 
We note that here $\Delta$ is calculated by optimizing over $\vec\lambda\oplus \vec{\lambda'}$ associated with the circuit
$\vec O\oplus \vec{O'}$, instead of merely optimizing over $ \vec{\lambda'}$.
Here $\oplus$ denotes the concatenation of the two vectors.
This procedure is repeated until the first-order error $\Delta$ is less than a threshold.

\begin{figure}[t]
 \includegraphics[width=.85\linewidth]{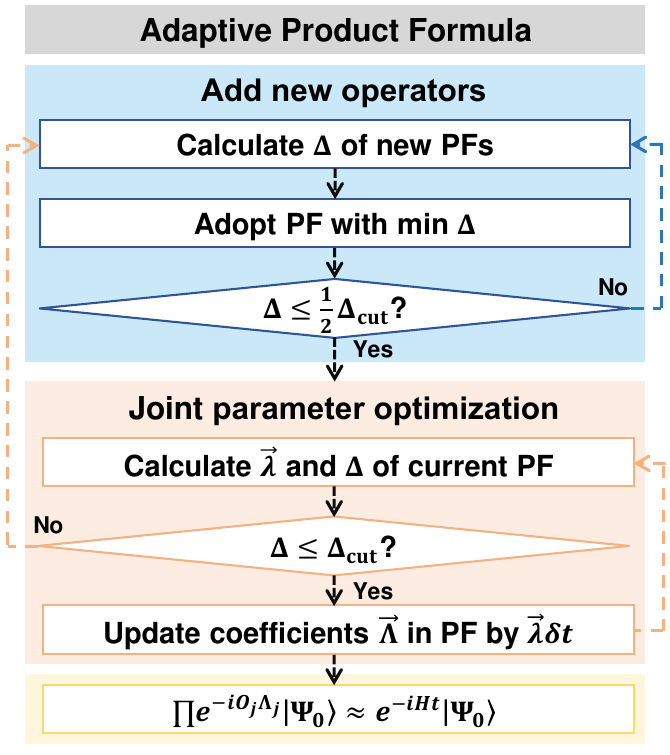}
 \caption{ 
 Scheme of the adaptive product formula (PF) method described in \autoref{alg:joint}. We adaptively construct the time evolution circuit of quantum states by finding the quantum operators and parameters that minimize the first-order error $\Delta$ at each step. 
 }\label{Fig:1}
\end{figure}

A major advantage of the new strategy is that, when the old parameters are allowed to change, we may only need to add much fewer new operators to ensure $\Delta$ is below the threshold. This is because we only add gates when $\Delta>\dcut$ and in the extreme case when optimizing over the existing parameters $\vec\lambda$ directly makes $\Delta\le\dcut$, no additional gates are needed to proceed the evolution.
In addition, we design that every time we construct the circuit, we add new operators until $\Delta\leq \dcut/2$, so that the following few evolution steps could be free from construction. We summarize the refined algorithm in \autoref{alg:joint} and \autoref{Fig:1}.

\begin{algorithm}[h]
\caption{Adaptive product formula}
\label{alg:joint}

1. Set $\dcut$. Input the initial state $\ket{\Psi_0}$ and Hamiltonian $H=\sum_{j=1}^{L} a_j P_j$;

2. (Joint  parameter optimization) Calculate optimal $\vec\lambda$ and $\Delta$ of the current circuit $G(\vec O,\vec\Lambda+\vec\lambda\delta t)\ket{\Psi_0}$ at time $t$. If $\Delta>\dcut$, go to the step 3. Otherwise, set $\vec\Lambda \rightarrow \vec\Lambda+\vec\lambda\delta t$ and continue step 2 with $t\rightarrow t+\delta t$; Terminate when $t=T$;

3. (Add new operators) 
\begin{enumerate}[(a)]
\item For every Pauli word $P_k$ in $H$, calculate $\Delta$ of the circuit 
$G(\vec O\oplus P_k,(\vec\Lambda+\vec\lambda\delta t)\oplus \lambda'\delta t)\ket{\Psi_0}$.
\item With $O^{(\min)}$ being the Pauli word giving the smallest $\Delta$ in  step (3a), add $O^{(\min)}$ to the end of the product formula as $\vec{O}\rightarrow \vec{O}\oplus O^{(\min)}$ and $\vec{\Lambda}\rightarrow \vec{\Lambda}\oplus 0$;
\item If we have $\Delta \leq \dcut/2$, stop adding operators and go to step 2. Otherwise, go to step (3a).
\end{enumerate}
\end{algorithm}

We prove that the ``add new operators'' procedure of \autoref{alg:joint} also satisfies the same properties in \autoref{thm:n_iter_bound}, and we refer to \cite{NoteX} for the proof. 
Compared to the conventional product formula method, which applies a deterministic sequence of gates, our method provides a circuit-growth strategy that optimizes the quantum circuits with a much lower gate count. The asymptotic worst error bound of our method is $\mc O(\dcut  T)$ as proven in \cite{NoteX}. 

\begin{figure*}[t]
	\includegraphics[width=\linewidth]{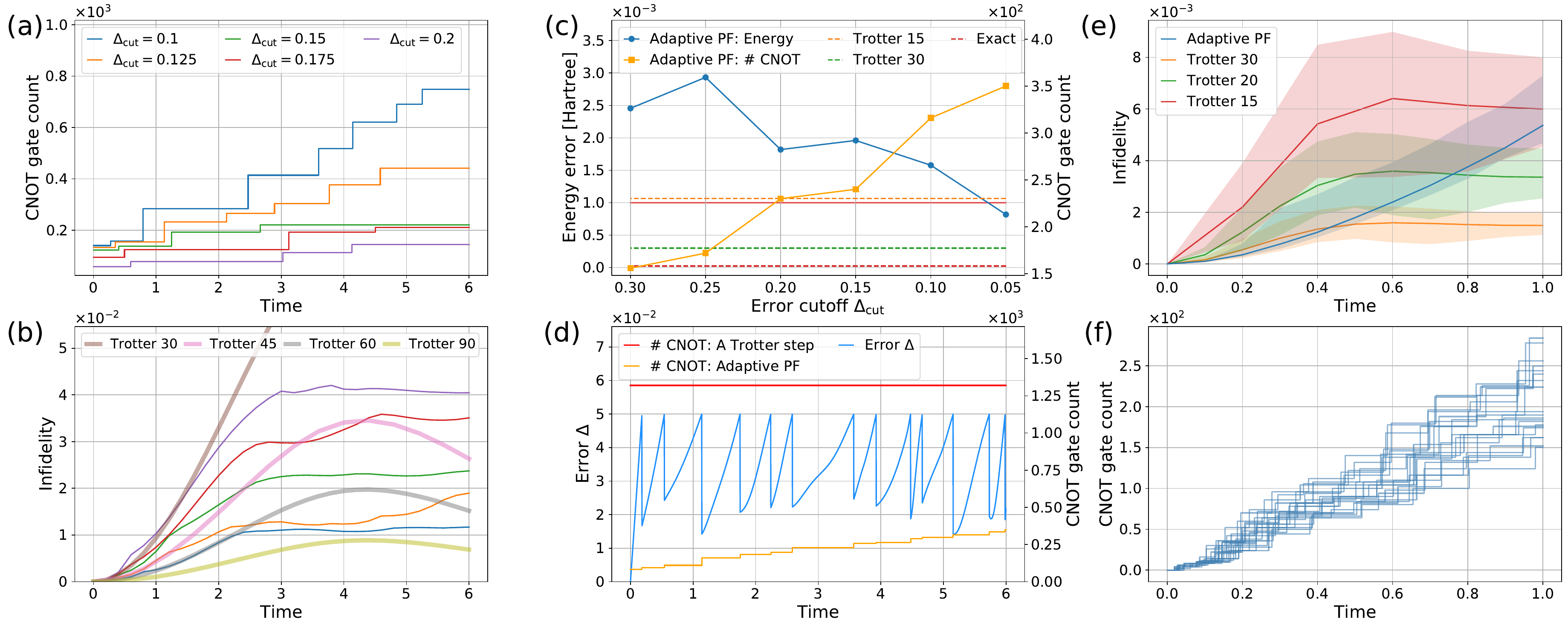}
\caption{ Simulation results of the adaptive product formula (PF) method. ``Trotter $n$''   specifies how many Trotter steps are used for evolving to $T$, i.e., with a step size $T/n$.
(a) CNOT gate counts of adaptive PF runs on $\mathrm{H_2O}$. 
The counts increase during the run as more CNOT gates are added to the circuit.
(b) Fidelities of adaptive PF on $\mathrm{H_2O}$ with comparison to Trotter methods to achieve $T=6$ (This figure also uses the legend of (a)). In the run of $\dcut=0.2$ ($144$ CNOTs used), the fidelity achieved is better than the first-order Trotter with $30$ steps ($1.6\times 10^5$ CNOTs used).
(c) Ground state energy of $\mathrm{H_4}$ obtained by the quantum Krylov algorithm with different time evolution methods. The CNOT counts of adaptive PF runs are also presented. Adaptive PF, first-order Trotter and exact evolution are compared. To achieve the chemical accuracy, more than $m=15$ time steps (for evolving to $T=mt=6$) are required for the first-order Trotter method ($1.98\times 10^4$ CNOTs used). In contrast, adaptive PF only uses $350$ CNOT gates.
(d) Trend of the error $\Delta$ and CNOT count in one run of adaptive PF on $\mathrm{H_4}$ with $\dcut=0.05$. The error $\Delta$ is controlled below $\dcut$ in the evolution. The CNOT gates required by adaptive PF are much fewer than that in one step of the first-order Trotter.
(e) Fidelities of adaptive PF on $20$ instances of random transverse field Ising model (TFIM). The solid lines represent the average fidelity. The shadows around the solid lines show the range of fidelities of the instances. Adaptive PF achieves similar accuracy as first-order Trotter.
(f) CNOT gate counts of adaptive PF on the instances of random TFIM. About $200$ CNOTs are needed to evolve to $T=1$ on average. In contrast, first-order Trotterization with similar accuracy needs about $1.98\times 10^3$ CNOTs.
}
\label{fig:numerics}
\end{figure*}

\emph{\textbf{Numerical Examples}.---}
Here, we compare the performance (specifically the number of CNOT gates~\footnote{The CNOT gate count is computed by assuming an explicit decomposition of $e^{-iPx}$ using $2n_q-2$ CNOT gates, where the Pauli word $P$ applies nontrivially on $n_q$ qubits.}) of our adaptive PF method to the first-order Suzuki-Trotter product formula for two molecular systems and a spin system. We choose a step size $\delta t$ of $2\times 10^{-3}$ for our method.

We first consider the $\mathrm{H_2O}$ molecule in equilibrium geometry with the 6-31g basis set. The Bravyi-Kitaev transformation is used for fermion-to-qubit mapping~\cite{bravyi2002fermionic} and the active-space Hamiltonian~\footnote{The active space is chosen to be 6 orbitals (A1:3, B1:1, B2:2) with 6 electrons.} contains $12$ qubits with $550$ Pauli terms and coefficient weight $\|a\|_1 = 16.690$ (trivial term excluded).
We test our method with different thresholds $\dcut$ and use first-order Trotter method as references.
We start from the Hartree-Fock state and evolve for a total time of $T=6$.
\autoref{fig:numerics}(a)(b) show the CNOT gate count and fidelity for each $\dcut$.
To achieve same accuracy, CNOT gates required by adaptive PF is much fewer than the Trotter method. With $\dcut=0.2$, the adaptive PF uses $144$ CNOTs gates whereas $1.6\times 10^5$ CNOT gates ($30$ Trotter steps) are needed for the first-order Trotter method to achieve similar fidelity.

Next, we simulate the time evolution of $\mathrm{H_4}$, a chain of hydrogen atoms spaced by $1.5\mathrm{\AA}$ starting from its Hartree-Fock state, and apply it in the quantum Krylov algorithm~\cite{stair2020multireference}, which projects and solves the ground energy of a system in the subspace spanned by $\left\{e^{iHnt}\ket{\Phi_0}\big|n\in 0,1,\dots,m \right\}$.  
The Hamiltonian~\footnote{We choose the STO-3g basis and use the Bravyi-Kitaev transformation.} involves $8$ qubits and $184$ Pauli words, with $\|a\|_1 = 5.654$.
The result is shown in \autoref{fig:numerics}(c).
When $\dcut=0.05, t=0.4$ and $m=15$, our method uses $350$ CNOT gates to simulate the whole time evolution and the chemical accuracy $10^{-3}$ Hartree is achieved. In contrast, the first-order Trotter with $1$ Trotter steps in every time interval $t$ requires totally $15\times 1320=1.98\times 10^4$ CNOT gates, and fails to achieve chemical accuracy. 
\autoref{fig:numerics}(d) shows how the distance $\Delta$ varies in the simulation. Every time $\Delta$ exceeds $\dcut$, an adaptive construction is carried out to reduce it to $\dcut/2$.

Finally, we simulate an $12$ qubit random transverse field Ising model (TFIM) with Hamiltonian $H=\sum_{i,j}^{n} w_{i,j} Z_i Z_j + \sum_k^{n} h_k X_k$, and the parameters $w_{i,j}$ and $h_k$ are uniformly sampled from $[-1,1]$ for every $i>j$ and $k$~\footnote{There are $n(n-1)/2+n=78$ terms in the Hamiltonian and the coefficients are then normalized so that the sum of their absolute value is $78\times 0.5=39.0$.}.  The importance of TFIM  is from its universality to encode computational problems such as MAXCUT~\cite{crooks2018performance} and it is widely used in quantum approximate optimization algorithms (QAOA)~\cite{farhi2014quantum}.
We sampled 20 instances
, where each instance has initial state $\ket{0}^{\otimes 12}$, $T=1$, and $\dcut=0.2$. As shown in \autoref{fig:numerics}(e), adaptive PF achieves a better accuracy compared to the first-order Trotter method with $15$ Trotter steps, which requires $1980$ CNOT gates for simulating the whole evolution.
In contrast, as shown in \autoref{fig:numerics}(f), adaptive PF only requires about $200$ CNOT gates, which is about two orders reduction.

\emph{\textbf{Shot noise analysis}.---}
We further discuss how  statistical error due to a finite number of measurements affects the estimation of $\Delta^2$, which we denote as $\Delta'^2$. 
Since $A$ and $C$ are not estimated accurately in the presence of shot noise,  directly solving $A\vec{\lambda}=C$ may result in  $\vec{\lambda}$ whose $\Delta'^2$ has a large variance and hence has a large fluctuation around the true value.
A solution to this problem is to find  $\vec{\lambda}$ which keeps both the mean value of $\Delta'^2$ and its variance $\var[\Delta'^2]$ small.
We consider the random variable $\Delta'^2(\vec{\lambda})$, which represents the estimation of $\Delta^2$ for certain $\vec{\lambda}$, whereas $\vec{\lambda}$ is not determined by solving the aforementioned linear equation.
As $\Delta'^2(\vec{\lambda})$ is unbiased, the variance $\var[\Delta'^2(\vec{\lambda})]$ characterizes its deviation from the true value.

When the entries of $A$ and $C$ are measured by $l_1$ sampling~\footnote{Here, $\braket{H^2}$ is assumed to be accurately estimated and the variance of $\braket{H^2}$ is negligible because $\braket{H^2}$ is invariant under the ideal time evolution $e^{-iHt}$.
}, with an optimized distribution of measurements, we have
\begin{align}
\label{equ:measure-var}
    \var[\Delta'^2(\vec{\lambda})] 
    &\leq (\|\vec{\lambda}\|_1^2+2\|a\|_1\|\vec{\lambda}\|_1)^2{n}^{-1}_M,
\end{align}
where $n_M$ is the number of measurements and $\|\cdot\|_1$ stands for the $l_1$ norm. 
The above bound indicates that the variance of $\Delta'^2(\vec{\lambda})$ can be reduced by choosing $\vec{\lambda}$ with a smaller norm.
Therefore, we propose to determine $\vec{\lambda}$ by minimizing a regularized objective $\Delta_{\mathrm{reg}}'^2(\vec{\lambda}) = \Delta'^2(\vec{\lambda})+\beta \|\vec{\lambda}\|_1$ \footnote{Here, $\beta$ is a positive number to be set. Note that  $\Delta_{\mathrm{reg}}'^2(\vec{\lambda})$ can be minimized by solving linear equations. See Supplemental Materials~\cite{NoteX} for detail.}, so that $\vec{\lambda}$ that makes the estimated $\Delta(\vec{\lambda})$ and the estimation error both small can be found.
We further address that an acceptable $\vec{\lambda}$ such that $\Delta\leq\dcut/2$ and $\|\vec{\lambda}\|_1\leq\|a\|_1$ always exists by adding new operators to the product formula, which guarantees a bounded variance (see Sec. III~C in \cite{NoteX}).
The measurement cost can be bounded even with a large circuit as \autoref{equ:measure-var} is independent of the circuit size. 
However, while adding more operators reduces the statistical error, it results in deeper circuits and thus there is a trade-off concerning hardware noise.

\emph{\textbf{Conclusion \& Outlooks}.---}
In this work, we propose an adaptive protocol to learn an optimized time evolution circuit with a fixed unknown input quantum state. 
Our protocol significantly suppresses the depth of circuits (by about two orders) compared to those of first-order product formula methods.
Our algorithm could be useful in methods that require Hamiltonian simulation as a subroutine, such as adiabatic quantum state preparation~\cite{RevModPhys.90.015002,babbush2014adiabatic} and finding eigenvalues of Hamiltonians~\cite{stair2020multireference,motta2020determining, parrish2019quantum,somma2019quantum,kyriienko2020quantum,endo2020variational,2021arXiv210915304Z}.
We show that advanced measurement schemes~\cite{verteletskyi2020measurement,wu2021overlapped,kandala2017hardware,huang2020predicting,hadfield2020measurements,bonet2020nearly,zhang2021experimental} can be applied to reduce the measurements (see 
Sec.~III~B in \cite{NoteX}).
The simulation accuracy in the presence of device errors could be further improved by combining with quantum error mitigation methods \cite{li2017efficient,endo2018practical,temme2017error,endo2020hybrid,PhysRevA.95.042308,strikis2020learning,bravyi2020mitigating,sun2020mitigating2,dumitrescu2018cloud,otten2019accounting,mcclean2020decoding}. 
Meanwhile, variational quantum algorithms have been developed to approximate the evolution using a fixed parameterized quantum circuit~\cite{li2017efficient,mcardle2019variational,cerezo2020variational,endo2020variational,yuan2019theory}.
However, the simulation accuracy could be limited with an imperfect circuit approximation and shot noise.
Our work addresses this challenge by adaptively learning the circuit that minimizes the simulation error.

\begin{acknowledgments}
J.S. thanks Suguru Endo and Z.Z. thanks Renke Huang and Bin Cheng for helpful related discussions.
M.Y. was supported by the Natural Science Foundation of Guangdong Province (Grant NO.2017B030308003), the Key R\&D Program of Guangdong province (Grant NO. 2018B030326001), the Science, Technology and Innovation Commission of Shenzhen Municipality (Grant NO. JCYJ20170412152620376 and NO. JCYJ20170817105046702 and NO. KYTDPT20181011104202253), National Natural Science Foundation of China (Grant NO. 11875160 and NO. U1801661), the Economy, Trade and Information Commission of Shenzhen Municipality (Grant NO.201901161512), Guangdong Provincial Key Laboratory (Grant NO. 2019B121203002), the Open Project of Shenzhen Institute of Quantum Science and Engineering (Grant No.~SIQSE202008), the National Natural Science Foundation of China Grant No.~12175003.
The quantum circuit simulation is done by Qulacs~\cite{suzuki2020qulacs}. PySCF~\cite{sun2018pyscf,10.1002/jcc.23981}. OpenFermion~\cite{mcclean2020openfermion} are used to generate the molecular Hamiltonian and transform them into qubit Hamiltonians. 
\\

\noindent\emph{Note added.}---
Another related work has appeared around the same time when our paper was posted on arXiv~\cite{yao2020adaptive}.
This work discussed adaptive variational quantum algorithms for simulating the quantum dynamics similar to Protocol 2 in our work. Comparison of the theory is presented in \cite{NoteX}.
\end{acknowledgments}
 
\bibliographystyle{apsrev4-2}
\bibliography{main}

\appendix

\widetext

\section{Adaptive product formula}

In this section, we discuss how to adaptively find  a short  product  formula  for Hamiltonian simulation from  a fixed  input  quantum  state. We first analyze the algorithmic errors for the adaptive product formula method, and based on that, we discuss the resource requirements to achieve a given simulation accuracy. Finally, we discuss the relations to prior works and the quantum resources for implementation.
    
\subsection{Adaptive product formula (single step) }
\label{SM:single-step}	
    In the single-step protocol, we approximate the exact time-evolved state $e^{-iH \delta t}\ket{\Psi(t)}$  by  \begin{equation}
 \ket{\Psi(t+\delta t)}=\prod_j e^{-iO_j\lambda_j \delta t}\ket{\Psi(t)}
\label{eqn:dis1},
\end{equation}
where a tuple $\vec O= \{O_j\}$ at each step is chosen from the Hamiltonian terms and the coefficients $\vec \lambda = \{\lambda_j\}$ are real. 
The error for the approximation at each finite time step can be described by the Euclidean distance between the exact time-evolved state and the approximated state
\begin{equation}
\tilde \varepsilon = \bigg\| e^{-iH \delta t}\ket{\Psi(t)}-\prod_j e^{-iO_j\lambda_j \delta t}\ket{\Psi(t)} \bigg\|,
\label{eqn:suppdis}
\end{equation}
where $\|\ket{\psi}\|=\sqrt{\braket{\psi|\psi}}$ represents the Frobenius norm.

Using Taylor expansion $e^{-iH\delta t}= \sum_{k=0}^{\infty}\frac{\delta t ^k}{k!} (-iH)^k$ and $\prod_j e^{-iO_j\lambda_j \delta t}=\sum_{k=0}^{\infty}\sum_{j_1,j_2,...,j_k}\frac{ (-i)^k \delta t ^k}{k!}  \prod_{k}(\lambda_{j_k}   O_{j_k}) $
, the algorithmic error can be expressed by
\begin{equation}
\tilde\varepsilon  = \sqrt{\Delta^2 \delta t^2 +\mathcal{O}(\delta t^3)}\leq { \Delta  \delta t +\mathcal{O}(\delta t^{3/2})}
\end{equation}
with the first-order error $\Delta$ being 
\begin{equation}
\begin{aligned}
\Delta^2 =&\braket{ H^2 }+\sum_{j j'}  A_{jj'} {\lambda}_j {\lambda}_{j'} -2\sum_{j } C_j {\lambda}_j.     
\end{aligned}
\end{equation}
Here, we define $A_{jj'}=\text{Re} \left(\braket{\Psi(t)|O_jO_{j'}|\Psi(t)}\right)$,  $C_j=\text{Re} \left( \braket{\Psi(t)|HO_j|\Psi(t)}\right) $ and $\braket{ H^2 }=\braket{\Psi(t)|H^2|\Psi(t)}$.

With a given $\vec O$, $\Delta^2$ is a quadratic function with respect to $\vec \lambda$, whose minimum can be obtained at the stationary point where 
$\frac{\partial}{\partial \lambda_j} \Delta = 0$ for all $j$. This is equivalent to $
\sum_{j'} A_{jj'} \lambda_{j'} =  C_j
$, or
$A \vec\lambda =C$ (in a vector form). Therefore, the coefficients $\vec \lambda$ can be determined by solving the linear equation,  either by applying the inverse matrix $A^{-1}$ or by an iterative algorithm. We note that in practice when the solution $\vec\lambda$ has a large vector norm, one may need to modify the linear system so that a solution with a smaller norm can be obtained. This is important because the norm of $\vec{\lambda}$ is directly related to the finite time step error (see \autoref{SM:error-finite-step}) and measurement cost (see \autoref{SM:shotnoiseanalysis}).
The accuracy of time evolution at each time step can be bounded by error $\mathcal{O}(\Delta \delta t)$. Therefore, $\Delta$ (or $\Delta^2$) serves as a handy measurable quantifier to estimate the quality of the time evolution with a choice of $\vec O$. 
To construct a circuit (which can be represented by $\vec O$) that approximates the exact evolution with a short circuit depth and a guaranteed simulation accuracy, we can just add a new Pauli word to $\vec O$ and see the change of $\Delta$. In the single-step strategy, we iterate over all the Pauli words in the Hamiltonian and add the Pauli word that gives the lowest $\Delta$ to $\vec O$.
As the total error can be bounded by $\varepsilon_{\mathrm{total}}\leq T \max\Delta $ given small $\delta t$, we can set a threshold $\dcut=\varepsilon_{\mathrm{total}}/T$, and if $\Delta>\dcut$ for the new circuit, we repeat to add new operators until $\Delta<\dcut$. This adaptive operator-adding (circuit construction) procedure can suppress the algorithmic error in time evolution to any given $\varepsilon_{\mathrm{total}}$ with proper $\dcut$ and $\delta t$.

\subsection{Adaptive product formula (jointly optimized)}
 \label{SM:jointPF}

First, let us define $\prod_{j}e^{-iO_j\Lambda_j}$ to be $G(\vec O,\vec\Lambda)$ for conciseness.
Suppose we want to construct a circuit at time $t$.
The single-step strategy is to consider the approximation  by $ G(\vec{O_t'}, \vec{\lambda_t'} \delta t)G(\vec {O}_t,\vec{\Lambda}_t) \ket{\Psi_0}  $, in which we fix the original circuit block $G(\vec O_t,\vec{\Lambda}_t)$ at time $t$  and optimize over the newly added circuit block $ G(\vec{O_t'}, \vec{\lambda_t'} \delta t)$. 
However, since our objective is to get the time-evolved state from $\ket{\Psi_0}$, rather than a general unitary for the time evolution, we may only need a much shorter circuit to effectively implement the state evolution to time $t$. In this section, we show how to jointly optimize the parameters and implement the state evolution up to a given simulation error.

In the jointly-optimized strategy, the exact time-evolved state $e^{-iH\delta t}G(\vec O_t,\vec\Lambda_t)\ket{\Psi_0} $  is approximated by 
\begin{equation}
G(\vec{O_t'},\vec{\lambda_t'}\delta t)G(\vec O_t,\vec\Lambda_t+{\vec\lambda_t} \delta t)\ket{\Psi_0}) = G(\vec O_t\oplus\vec{O_t'},(\vec\Lambda_t+{ \vec\lambda_t}\delta t)\oplus (\vec{\lambda_t'}\delta t))\ket{\Psi_0},
\end{equation}
where we used that 
\begin{equation}
\label{equ:rule-for-G-1}
G(\vec{O}_1,\vec{\Lambda}_1)G(\vec{O}_2,\vec{\Lambda}_2) = G(\vec{O}_1\oplus\vec{O}_2,\vec{\Lambda}_1\oplus\vec{\Lambda}_2).
\end{equation}

After this step at $t$, we update the operators and parameters by $\vec{O}_{t+\delta t} = \vec O_t\oplus\vec{O_t'}$ and $\vec{\Lambda}_{t+\delta t} = (\vec\Lambda_t+{ \vec\lambda}\delta t)\oplus (\vec{\lambda_t'}\delta t)$, respectively. The total error for the approximation at time $t+\delta t$ can be described by the Euclidean distance of the exact time-evolved state and the approximated state, $\varepsilon_{\mathrm{total}} =  D(e^{-iH(t+\delta t)}\ket{\Psi_0}  ,G(\vec{O_t'},\vec{\lambda_t'}\delta t)G(\vec {O}_t,\vec{\Lambda}_t+{\vec\lambda_t} \delta t)\ket{\Psi_0})$  where  $D(\rho,\sigma)=\|\rho-\sigma\|$ represents the Euclidean distance.
Using the triangle inequality,  the total error is upper bounded by 
\begin{equation}\label{Eq:tridistance}
\begin{aligned}
\varepsilon_{\mathrm{total}} & \leq D(e^{-iH(t+\delta t)}\ket{\Psi_0}  , e^{-iH\delta t}G(\vec O_t,\vec\Lambda_t )\ket{\Psi_0})+ D(e^{-iH\delta t}G(\vec O_t,\vec\Lambda_t )\ket{\Psi_0},G(\vec{O_t'},\vec{\lambda_t'}\delta t)G(\vec O_t,\vec\Lambda_t+{\vec\lambda_t} \delta t )\ket{\Psi_0}) \\
& \leq  \sum_{t \in \text{time steps}}  D(e^{-iH\delta t}G(\vec O_t,\vec\Lambda_t )\ket{\Psi_0},G(\vec{O_t'},\vec{\lambda_t'}\delta t)G(\vec O_t,\vec\Lambda_t+{\vec\lambda_t}\delta t)\ket{\Psi_0}).
\end{aligned}
\end{equation} 
Here, we used the invariance of distances under unitary transformation and the summation in the second line is made over all the $t$ at which a time step is made. We define
\begin{align}\label{Eq:stepdistance}
    \tilde \varepsilon_t&=D(e^{-iH\delta t}G(\vec O_t,\vec\Lambda_t )\ket{\Psi_0},G(\vec{O_t'},\vec{\lambda_t'}\delta t)G(\vec O_t,\vec\Lambda_t+ {\vec\lambda_t} \delta t)\ket{\Psi_0})
\end{align}
which characterizes the approximation error with the circuit  $G(\vec{O_t'},\vec 0)G(\vec O_t,\vec\Lambda_t)$. This definition gives $\varepsilon_{\mathrm{total}} \leq \sum_{t \in \text{time steps}} \tilde \varepsilon_t$.
Notice \autoref{equ:rule-for-G-1} and the relation $G(\vec{O}_1\oplus\vec{O}_2, \vec{\Lambda}_1\oplus\vec{0})=G(\vec{O}_1, \vec{\Lambda}_1)$, we have
\begin{align}
    \tilde  \varepsilon_t = D(e^{-iH\delta t}G(\vec O_t\oplus \vec{O_t'},\vec\Lambda_t\oplus \vec 0 )\ket{\Psi_0},G(\vec O_t\oplus \vec{O_t'},(\vec\Lambda_t\oplus \vec 0)
    + ({\vec\lambda_t\oplus \vec{\lambda_t'}}) \delta t)\ket{\Psi_0}).
\end{align}
By setting $\vec O\leftarrow \vec O_t\oplus\vec {O_t'}$, $\vec\Lambda \leftarrow \vec\Lambda_t\oplus \vec 0$ and ${\vec\lambda} \leftarrow ({\vec{\lambda_t}}\oplus \vec{\lambda_t'})$, we have the form
\begin{equation}
\label{equ:simplified-form-epsilon}
\tilde  \varepsilon = D(e^{-iH\delta t}G(\vec O,\vec\Lambda )\ket{\Psi_0},G(\vec O,\vec\Lambda+ {\vec\lambda} \delta t)\ket{\Psi_0}).
\end{equation}
In the following, we give a first-order approximation of $\tilde  \varepsilon$, so that $\tilde \varepsilon_t$ can be approximated. To start with, we rewrite $\tilde  \varepsilon$ by
\begin{equation}\label{equ:error:expanded}
\begin{split}
     \tilde \varepsilon&=\sqrt{\|e^{-iH\delta t}G(\vec O,\vec\Lambda )\ket{\Psi_0}-G(\vec O,\vec\Lambda+{\vec\lambda}\delta t)\ket{\Psi_0}\|^2}.\\
\end{split}
\end{equation}
Denote the differential operator on $\Lambda_j$ as $\partial_j\equiv \partial/\partial\Lambda_j$ and $\partial_{j j'} \equiv \partial_{j}\partial_{j'}$. We can expand $G(\vec O,\vec\Lambda+{\vec\lambda}\delta t)$ as 
\begin{equation}
   G(\vec O,\vec\Lambda+{\vec\lambda}\delta t) =G(\vec O,\vec\Lambda ) +  \sum_j \lambda_j \partial_j  G(\vec O,\vec\Lambda) \delta t + \frac{1}{2}\sum_{jj'} \lambda_j \lambda_{j'} \partial_j \partial_{j'} G(\vec O,\vec\Lambda) \delta t^2 + \mathcal{O}(\delta t^3).
\end{equation}
Therefore, we have
\begin{align}
&e^{-iH\delta t}G(\vec O,\vec\Lambda )-G(\vec O,\vec\Lambda+{\vec\lambda}\delta t)\\
=&  -iHG(\vec O,\vec\Lambda )\delta t -  \sum_j \lambda_j \partial_j  G(\vec O,\vec\Lambda) \delta t - \frac{1}{2}H^2  G(\vec O,\vec\Lambda ) \delta t^2 - \frac{1}{2}\sum_{jj'} \lambda_j \lambda_{j'} \partial_{jj'}^2 G(\vec O,\vec\Lambda) \delta t^2 + \mathcal{O}(\delta t^3) 
\end{align}
and
\begin{align}
\label{equ:delta-1-2}
&\|(e^{-iH\delta t}G(\vec O,\vec\Lambda )-G(\vec O,\vec\Lambda+{\vec\lambda}\delta t))\ket{\Psi_0}\|^2\\
=&  \|-( iH +  \sum_j \lambda_j \partial_j )  G(\vec O,\vec\Lambda) \ket{\Psi_0}\delta t - \frac{1}{2}(H^2 + \sum_{jj'} \lambda_j \lambda_{j'} \partial_{jj'}^2 )G(\vec O,\vec\Lambda) \ket{\Psi_0} \delta t^2 + \mathcal{O}(\delta t^3) \|^2 \\
= & \|\ket{\psi_1}\delta t + \ket{\psi_2}\delta t^2+ \mathcal{O}(\delta t^3) \|^2 = \ketbra{\psi_1}{\psi_1} \delta t^2 + 2\text{Re} \left( \ketbra{\psi_1}{\psi_2} \right)  \delta t^3 + \mathcal{O}(\delta t^4),
\end{align}
where 
\begin{align}
    \ket{\psi_1}  \equiv -( iH +  \sum_j \lambda_j \partial_j)  G(\vec O,\vec\Lambda) \ket{\Psi_0},~~~     \ket{\psi_2}  \equiv  - \frac{1}{2}(H^2 + \sum_{jj'} \lambda_j \lambda_{j'} \partial_{jj'}^2 ) G(\vec O,\vec\Lambda)\ket{\Psi_0}.
\end{align}

Substituting the above expansion into \autoref{equ:error:expanded},  we have
\begin{equation} \label{equ:second:Delta}
		\tilde\varepsilon  = \sqrt{\Delta^2 \delta t^2 + \Delta_2^2 \delta t^3 +\mathcal{O}(\delta t^4)}  \leq \Delta \delta t+ \Delta_2 {\delta t}^{3/2} +\mathcal{O}(\delta t^2).
	\end{equation}

The first-order order error $\Delta$ can be expressed as 
\begin{equation}
\begin{aligned}
\Delta^2 = \ketbra{\psi_1}{\psi_1} =\braket{H^2}  + \sum_{jj'} A_{jj'} \lambda_j \lambda_{j'}  - 2 \sum_j C_j \lambda_j,
\label{equ:SM_Delta}
\end{aligned}
\end{equation} 
where we denote 
\begin{equation}
\begin{aligned}
\label{equ:A-C-define}
A_{jj'}= \operatorname{Re}\left( \bra{\Psi_0}\partial_{j}G^{\dagger}(\vec O,\vec\Lambda )  \partial_{j'}G(\vec O,\vec\Lambda  )\ket{\Psi_0} \right),~~~
C_j = \operatorname{Im} \left( \braket{\Psi_0 |\partial_{j} G^{\dagger}(\vec O,\vec\Lambda ) H  G(\vec O,\vec\Lambda)  |\Psi_0} \right),
\end{aligned}
\end{equation}
and 
\begin{equation}
\braket{ H^2 }=\braket{\Psi_0|G^{\dagger}(\vec O,\vec\Lambda )H^2G(\vec O,\vec\Lambda ) |\Psi_0}.
\end{equation}
Finally, we can represent the total error by

\begin{align}
\label{equ:total-error}
	\varepsilon_{\mathrm{total}} \leq \sum_{t \in \text{time steps}} \tilde \varepsilon_t \leq \Delta^{(\mathrm{max})} T+\Delta_2^{(\mathrm{max})} \sqrt{\delta t} T.
\end{align}
Note that, the same as the single-step strategy, to minimize $\Delta$, the coefficients $\vec \lambda$ here can be determined by solving the linear equation as  $ \sum_{j'} A_{jj'} \lambda_{j'}=C_j$.
In \autoref{SM:measure-AC}, we show how the entries of $A$ and $C$ can be evaluated by quantum circuits.

One difference between the jointly-optimized strategy and the single-step strategy is that, when old parameters are allowed to change, $\Delta$ of the circuit may be lower than the cutoff $\dcut$ just by using the old circuit itself. In this case, adding new operators becomes unnecessary. Therefore, in the jointly-optimized strategy, we only add new operators when $\Delta>\dcut$, which could largely reduce the quantum gates required for certain simulation accuracy. In the main text, we proposed to add new operators until $\Delta\leq \dcut/2$ to avoid constructing the circuit too frequently.

\subsection{Error from the finite time step}
\label{SM:error-finite-step}
We next consider the error due to a finite time step, which is quantified by $\Delta_2$.  
According to \autoref{equ:total-error}, to suppress the total algorithmic error to $\varepsilon_{\mathrm{total}}$ with  $\Delta^{(\mathrm{max})} T=\varepsilon_{\mathrm{total}}/2$ and $\Delta_2^{(\mathrm{max})} \sqrt{\delta t} T=\varepsilon_{\mathrm{total}}/2$, we need to ensure $\Delta^{(\mathrm{max})} \leq \varepsilon_{\mathrm{total}}/(2T)$ and
set the time step $\delta t \leq \varepsilon_{\mathrm{total}}^2/(4\Delta_2^{2(\mathrm{max})} T^2)$. Hence, we require the number of steps to be $N=T/\delta t \geq 4 \Delta_2^{2(\mathrm{max} ) } T^3/ \varepsilon_{\mathrm{total}}^2$.

From \autoref{equ:delta-1-2}, the error due to a finite time step, $\Delta_2$, can be expanded as
\begin{align}
 \Delta_2^2 = &2 \operatorname{Re} \left( \ketbra{\psi_1}{\psi_2} \right)  \\
     =& -   \operatorname{Re} \left(\bra{\Psi_0}   \big(iG^{\dagger}(\vec O,\vec\Lambda)H  - \sum_j { {\lambda}_j \partial_jG^{\dagger}(\vec O,\vec\Lambda) }  \big)  \big(H^2 + \sum_{j j'}  {\lambda}_j  {\lambda}_{j'}  \partial^2_{jj'} \big)   G(\vec O,\vec\Lambda)\ket{\Psi_0} \right) \\
     = & \operatorname{Im}\left(   \bra{\Psi_0}  G^{\dagger}(\vec O,\vec\Lambda) H \sum_{j j'}  {\lambda}_j  {\lambda}_{j'}  \partial_{jj'}^2  G(\vec O,\vec\Lambda)\ket{\Psi_0} \right)
     + \operatorname{Re} \left( \bra{\Psi_0}  G^{\dagger}(\vec O,\vec\Lambda) H^2 \sum_j { {\lambda}_j \partial_j }  G(\vec O,\vec\Lambda)\ket{\Psi_0} \right) \label{equ:Delta22-sub-1} \\
     &  +  
     \operatorname{Re} \left( \bra{\Psi_0}  \sum_j { {\lambda}_j \partial_j } G^{\dagger}(\vec O,\vec\Lambda) \sum_{j' j''}  {\lambda}_{j'}  {\lambda}_{j''}  \partial_{j'j''}^2  G(\vec O,\vec\Lambda)\ket{\Psi_0} \right).\label{equ:Delta22-sub-2}
\end{align}
In the third equality, we used the fact that $\bra{\psi}H^3\ket{\psi}$ is always real.

In our case, where $\{O_j\}$ are Pauli operators, we have $\partial_j e^{-iO_j\Lambda_j} = -iO_j e^{-iO_j\Lambda_j}$, which is a unitary. This implies that $\partial_j G(\vec O,\vec\Lambda)$ and $\partial_{jj'} G(\vec O,\vec\Lambda)$ are also unitary. Therefore, $\partial_j G(\vec O,\vec\Lambda)\ket{\Psi_0}$ and $\partial_{jj'}^2 G(\vec O,\vec\Lambda) \ket{\Psi_0}$ can be treated as quantum states, and their norms are 1,
\begin{align}
    \Vert \partial_j G(\vec O,\vec\Lambda)\ket{\Psi_0} \Vert = 1, \quad
    \Vert \partial_{jj'}^2 G(\vec O,\vec\Lambda) \ket{\Psi_0} \Vert = 1.
\end{align}
Thus, we can simplify $\Delta_2$ by taking a norm on the summands in \autoref{equ:Delta22-sub-1} and \autoref{equ:Delta22-sub-2}, 
\begin{align}
\label{equ:delta2}
    \Delta_2^2 \leq & \left\Vert   \bra{\Psi_0}  G^{\dagger}(\vec O,\vec\Lambda) H \right\Vert \left( \sum_{j j'}  \left\vert  {\lambda}_j  {\lambda}_{j'}   \right\vert \left\Vert \partial_{jj'}^2  G(\vec O,\vec\Lambda)\ket{\Psi_0} \right\Vert \right)
     +  \left\Vert \bra{\Psi_0}  G^{\dagger}(\vec O,\vec\Lambda) H^2 \right\Vert \left( \sum_j \left\vert {\lambda}_j \right\vert \left\Vert \partial_j   G(\vec O,\vec\Lambda)\ket{\Psi_0} \right\Vert \right) \\
     &  +  
    \sum_{j j' j''}  \left\vert {\lambda}_j {\lambda}_{j'}  {\lambda}_{j''}  \right\vert\left\Vert \bra{\Psi_0}  \partial_j  G^{\dagger}(\vec O,\vec\Lambda) \partial_{j'j''}^2  G(\vec O,\vec\Lambda)\ket{\Psi_0} \right\Vert \\
    \leq & \left\Vert  H \right\Vert  \sum_{j j'}  \left\vert  {\lambda}_j  {\lambda}_{j'}   \right\vert
     +  \left\Vert H^2 \right\Vert\sum_j \left\vert {\lambda}_j \right\vert  +  
    \sum_{j j' j''}  \left\vert {\lambda}_j {\lambda}_{j'}  {\lambda}_{j''}  \right\vert
    =
    \left\Vert  H \right\Vert \Big( \sum_j \left\vert {\lambda}_j \right\vert \Big)^2
     +  \left\Vert H^2 \right\Vert \Big( \sum_j \left\vert {\lambda}_j \right\vert \Big)  +  
     \Big( \sum_j \left\vert {\lambda}_j \right\vert \Big)^3
    \\
    \leq & \Vert \vec\lambda \Vert_1^2 \Vert H \Vert +  \Vert \vec\lambda \Vert_1 \Vert H \Vert^2 + \Vert \vec\lambda \Vert_1^3,
\end{align}
where the norm $\Vert \cdot \Vert$ on $H$ and $H^2$ is the operator norm induced by the norm of quantum states, i.e.,
\begin{equation}
    \Vert H \Vert = \sup_{\ket{\psi}} \Vert H \ket{\psi} \Vert,
\end{equation} 
and the $l_1$ norm of the vector is defined by
\begin{equation}
    \Vert \vec\lambda \Vert_1 = \sum_j |\lambda_j|.
\end{equation}
It is also worth noting that $\Vert H \Vert \leq \Vert H \Vert_F = \sqrt{\sum_j a_j^2}$ given that $H=\sum_ja_jP_j$, where $\Vert \cdot \Vert_F$ is the Frobenius norm.
We can thus estimate the appropriate time step length during the run of the algorithm, provided the parameters $\vec \lambda$. In \autoref{SM:reduce-lambda-norm}, we will introduce a method to reduce $\Vert \vec\lambda \Vert_1$, so that $\Delta^2_2$ can be reduced.

\section{Analysis of the adaptive strategy}
\label{SM:thm} 

\subsection{Overview of the proof}
In this section, we validate our method by proving \autoref{thm:main-result} for the jointly-optimized strategy. The theorem in the main text for the single-step strategy can be seen as a trivial corollary of it.

The proof is based on the analysis of the tangent space of product formulas (See \autoref{def:tangent:pf}). In \autoref{thm:distance:error}, we prove that the first-order error is just the distance from the objective evolution direction to the tangent space of the current product formula. Then, in \autoref{thm:tangent:decompose}, we show how we can decompose the tangent space. After that, in \autoref{thm:two:orthogonal}, we present the condition 
for the error $\Delta$ (see rigorous definition in \autoref{equ-error}) to decrease when adding the new Pauli word to the product formula. Finally, by \autoref{thm:evolve:in:Ps} and \autoref{thm:evolve:in:T0}, we prove that the condition in \autoref{thm:two:orthogonal} is fulfilled in our method. 

Based on the lemmas, we prove our main result that guarantees the efficiency of our strategy. In \autoref{thm:must_decrease}, we prove that $\Delta$ must decrease in the ``add new operators'' procedure of Protocol 2 in the main text as long as $\Delta>0$. In \autoref{thm:most:add:once}, we prove that every Pauli word in the Hamiltonian can only be added once in a ``add new operators'' procedure, which guarantees the procedure terminates in finite steps. Finally, we summarize the main result in \autoref{thm:main-result}.

\subsection{Proofs}
We first define the error $\Delta$
as a function depending on $\vec{O}$, $\vec{\Lambda}$ and the initial state $\ket{\Psi_{0}}$ as

\begin{equation}
\Delta(\vec{O},\vec{\Lambda},\ket{\Psi_{0}})=\min_{\vec{\lambda}}\left\Vert \lim_{\delta t\rightarrow0}\frac{1}{\delta t}\left(e^{-iH\delta t}G(\vec{O},\vec{\Lambda})\ket{\Psi_{0}}-\prod_{j}e^{-iO_{j}(\Lambda_{j}+\lambda_{j}\delta t)}\ket{\Psi_{0}}\right)\right\Vert .\label{equ-error}
\end{equation}
From this definition, $\Delta(\vec{O},\vec{\Lambda},\ket{\Psi_{0}})$ is the minimized value of the first-order error $\Delta$ defined in \autoref{equ:SM_Delta}.

\newcommand{\tlim}{\delta t \rightarrow 0}
\newcommand{\limexpH}{\lim_{\tlim}\frac{1}{\delta t} e^{-iH\delta t} G(\vec{O},\vec{\Lambda})}
\newcommand{\twonorm}[1]{\left\Vert#1\right\Vert}

To study the minimum of the above difference, we define the tangent space of a product formula, which consists of all possible directions of change by updating its parameters. The minimum error can be then obtained by minimizing over the tangent space (See \autoref{thm:distance:error}).

\begin{definition}[Tangent space of PF] 
\label{def:tangent:pf}
Define the tangent space
$\mathcal{T}(\vec{O},\vec{\Lambda},\ket{\Psi_{0}})$ of a product formula
$(\vec{O},\vec{\Lambda})$ on state $\ket{\Psi_{0}}$ as 
\begin{equation}
\mathcal{T}(\vec{O},\vec{\Lambda},\ket{\Psi_{0}})=\{\lim_{\tlim}\frac{1}{\delta t}\prod_{j}e^{-iO_{j}(\Lambda_{j}+\lambda_{j}\delta t)}\ket{\Psi_{0}}|\lambda_{j}\in \mathbf{R}\}
\end{equation}
\end{definition}
We also define the set of Pauli words from the Hamiltonian, which  will be used frequently in the following proofs.
\begin{definition}[Pauli word pool from Hamiltonian]
Define $\mathbf{H}$ as the set that contains all the Pauli words that appear in the Pauli word decomposition of Hamiltonian $H=\sum_j a_j P_j$.
\end{definition}

We give of a brief summary of \autoref{thm:distance:error} and \autoref{thm:tangent:decompose}. 
The first lemma shows that $\Delta$ is just the distance between the direction of an ideal evolution and the tangent space. The second lemma shows that we can decompose a tangent space of a tuple $(\vec O,\vec \Lambda,\ket{\Psi_0})$ into a span of two smaller tangent spaces, given that one of the entries of $\vec\Lambda$ is zero.

\begin{lemma} 
\label{thm:distance:error} 
The error $\Delta(\vec{O},\vec{\Lambda},\ket{\Psi_{0}})$
is the distance between 
$\limexpH\ket{\Psi_{0}}$ and
$\mathcal{T}(\vec{O},\vec{\Lambda},\ket{\Psi_{0}})$.
\end{lemma}
\begin{proof}
The distance between $\limexpH\ket{\Psi_{0}}$ and $\mathcal{T}(\vec{O},\vec{\Lambda},\ket{\Psi_{0}})$
is defined as 
\begin{equation}
\min_{\vec{T}\in\mathcal{T}(\vec{O},\vec{\Lambda},\ket{\Psi_{0}})}\twonorm{\limexpH\ket{\Psi_{0}}-\vec{T}}
\end{equation}
which is equivalent to 
\begin{equation}
\min_{\vec{\lambda}}\twonorm{\limexpH\ket{\Psi_{0}}-\lim_{\tlim}\frac{1}{\delta t}\prod_{j}e^{-iO_{j}(\Lambda_{j}+\lambda_{j}\delta t)}\ket{\Psi_{0}}},
\end{equation}
and the definition in \autoref{equ-error}. 
\end{proof}
The following lemma indicates we can decompose the tangent space if an entry in $\vec\Lambda$ is $0$.  $\spacespan$ is defined as 
\begin{equation}
    \spacespan\{S_1,S_2\} = \{a_1\vec s_1+a_2\vec s_2|a_1,a_2\in \mathbf{R},\vec s_1\in S_1, \vec s_2\in S_2\},
\end{equation}
where $\mathbf{R}$ means the set of real numbers.
\begin{lemma}
\label{thm:tangent:decompose} 
\item
\begin{equation}
\mathcal{T}(\vec{O}\oplus P',\vec{\Lambda}\oplus0,\ket{\Psi_{0}})=\spacespan\{\mathcal{T}(\vec{O},\vec{\Lambda},\ket{\Psi_{0}}),\mathcal{T}(P',0,G(\vec{O},\vec{\Lambda})\ket{\Psi_{0}})\}
\end{equation}
\end{lemma} 

\begin{proof}
\item
\begin{equation}
\mathcal{T}(\vec{O}\oplus P',\vec{\Lambda}\oplus0,\ket{\Psi_{0}})=\{\lim_{\tlim}\frac{1}{\delta t}(e^{-iP'\lambda'\delta t}\prod_{j}e^{-iO_{j}(\Lambda_{j}+\lambda_{j}\delta t)}\ket{\Psi_{0}})|\lambda_{j}\in \mathbf{R},\lambda'\in \mathbf{R}\}
\end{equation}

Notice that every vector in $\mathcal{T}(\vec{O}\oplus P',\vec{\Lambda}\oplus0,\ket{\Psi_{0}})$
can be decomposed as 
\begin{equation}
\label{equ:space-decomp-lemma}
\begin{split} & \lim_{\tlim}\frac{1}{\delta t}(e^{-iP'\lambda'\delta t}\prod_{j}e^{-iO_{j}(\Lambda_{j}+\lambda_{j}\delta t)}\ket{\Psi_{0}})\\
= & \lim_{\tlim}\frac{1}{\delta t}e^{-iP'\lambda'\delta t}\lim_{\tlim}(\prod_{j}e^{-iO_{j}(\Lambda_{j}+\lambda_{j}\delta t)}\ket{\Psi_{0}})+\lim_{\tlim}e^{-iP'\lambda'\delta t}\lim_{\tlim}\frac{1}{\delta t}(\prod_{j}e^{-iO_{j}(\Lambda_{j}+\lambda_{j}\delta t)}\ket{\Psi_{0}})\\
= & \lim_{\tlim}\frac{1}{\delta t}e^{-iP'\lambda'\delta t}G(\vec{O},\vec{\Lambda})\ket{\Psi_{0}}+\lim_{\tlim}\frac{1}{\delta t}(\prod_{j}e^{-iO_{j}(\Lambda_{j}+\lambda_{j}\delta t)}\ket{\Psi_{0}})\\
\equiv & T_{P'}+T_{0},
\end{split}
\end{equation}
where $T_{0}\in\mathcal{T}(\vec{O},\vec{\Lambda},\ket{\Psi_{0}})$ and
$T_{P'}\in\mathcal{T}(P',0,G(\vec{O},\vec{\Lambda})\ket{\Psi_{0}})$. Therefore, every vector in $\mathcal{T}(\vec{O}\oplus P',\vec{\Lambda}\oplus0,\ket{\Psi_{0}})$ can be decomposed into a sum of two vectors from $\mathcal{T}(\vec{O},\vec{\Lambda},\ket{\Psi_{0}})$ and $\mathcal{T}(P',0,G(\vec{O},\vec{\Lambda})\ket{\Psi_{0}})$. Also, by tracing \autoref{equ:space-decomp-lemma} bottom-up, one can find that the sum of two arbitrary vectors from $\mathcal{T}(\vec{O},\vec{\Lambda},\ket{\Psi_{0}})$ and $\mathcal{T}(P',0,G(\vec{O},\vec{\Lambda})\ket{\Psi_{0}})$ is in $\mathcal{T}(\vec{O}\oplus P',\vec{\Lambda}\oplus0,\ket{\Psi_{0}})$. These two relations imply these two sets are the subsets of each other, and thus we prove $\mathcal{T}(\vec{O}\oplus P',\vec{\Lambda}\oplus0,\ket{\Psi_{0}})= \spacespan\{\mathcal{T}(\vec{O},\vec{\Lambda},\ket{\Psi_{0}}),\mathcal{T}(P',0,G(\vec{O},\vec{\Lambda})\ket{\Psi_{0}})\}$.
\end{proof}

In the following, we will show that, if appending a new Pauli word cannot expand the tangent space and decrease its distance to the objective vector $\limexpH\ket{\Psi_{0}}$, the \textit{new component} of the new tangent space with respect to the old space, which is the vectors in $\mathcal{T}(P',0,G(\vec{O},\vec{\Lambda})\ket{\Psi_{0}})$ that is orthogonal to $\mathcal{T}(\vec{O},\vec{\Lambda},\ket{\Psi_{0}})$, cannot have an overlap to $\limexpH\ket{\Psi_{0}}$.

With this lemma, we can show if appending a new Pauli word to the product formula cannot reduce the error $\Delta$, the new part that the new Pauli word brings to the tangent space must be orthogonal to the objective direction. This also indicates that, once the new part has an overlap with the objective direction, $\Delta$ is decreased.

\begin{lemma} \label{thm:two:orthogonal} If the distance $D$ from
$\limexpH\ket{\Psi_{0}}$ to $\mathcal{T}(\vec{O}\oplus P',\vec{\Lambda}\oplus0,\ket{\Psi_{0}})$
is the same as it to $\mathcal{T}(\vec{O},\vec{\Lambda},\ket{\Psi_{0}})$,
then for all the vector $T_{P'}$ in $\mathcal{T}(P',0,G(\vec{O},\vec{\Lambda})\ket{\Psi_{0}})$
that is orthogonal to $\mathcal{T}(\vec{O},\vec{\Lambda},\ket{\Psi_{0}})$,
$T_{P'}$ must be also orthogonal to $\limexpH\ket{\Psi_{0}}$. \end{lemma} 
\begin{proof}
For simplicity, we denote $\limexpH\ket{\Psi_{0}}$ by $T_H$ in this proof. Notice that by the definition of distance, there exists a vector $T_0$ in $\mathcal{T}(\vec{O},\vec{\Lambda},\ket{\Psi_{0}})$ such that 
\begin{equation}
D = \twonorm{T_0-T_H}.
\end{equation}
Easily, we can show that $T_0-T_H$ is orthogonal to $\mathcal{T}(\vec{O},\vec{\Lambda},\ket{\Psi_{0}})$.

Suppose  $T'$ is a vector in $\mathcal{T}(P',0,G(\vec{O},\vec{\Lambda})\ket{\Psi_{0}})$ 
and $T'$ is orthogonal to $\mathcal{T}(\vec{O},\vec{\Lambda},\ket{\Psi_{0}})$
but not orthogonal to $T_H$. Notice that $T_0\cdot T'=0$ ($T_0$ is a vector in $\mathcal{T}(\vec{O},\vec{\Lambda},\ket{\Psi_{0}})$), we have 
\begin{equation}
	\begin{split}
		& \twonorm{T_0+\lambda T'-T_H}^2 \\
		= & \twonorm{T_0-T_H}^2 +  \lambda^2 \twonorm{ T'}^2 + 2(T_0-T_H)\cdot \lambda T' \\
		= & D^2 + \lambda^2 \twonorm{T'}^2 - 2\lambda T_H\cdot T'.
	\end{split}
\end{equation}
Next, we prove 
\begin{equation}
    \lambda^2 \twonorm{T'}^2 - 2\lambda T_H\cdot T' = \twonorm{T'}^2\lambda(\lambda  - \frac{2T_H\cdot T'}{\twonorm{T'}^2})
\end{equation}
can be less than zero and therefore $\twonorm{T_0+\lambda T'-T_H}$ can be less than $D$. By a simple analysis of the zero points of the above second-degree polynomial in $\lambda$, we can see there exists $\lambda'$ such that
\begin{equation}
\lambda^{'2} \twonorm{T'}^2 - 2\lambda' T_H\cdot T' =\twonorm{T'}^2\lambda'(\lambda'  - \frac{2T_H\cdot T'}{\twonorm{T'}^2}) < 0,
\end{equation}
as long as $T_H\cdot T'\neq 0$ (which has been assumed). Therefore, there is a vector $T_0+\lambda'T'$ such that 
\begin{equation}
\twonorm{T_0+\lambda' T'-T_H} < \twonorm{T_0-T_H}.
\end{equation}
Because $T_0 + \lambda' T' \in \spacespan\{\mathcal{T}(\vec{O},\vec{\Lambda},\ket{\Psi_{0}}),\mathcal{T}(P',0,G(\vec{O},\vec{\Lambda})\ket{\Psi_{0}})\} = \mathcal{T}(\vec{O}\oplus P',\vec{\Lambda}\oplus0,\ket{\Psi_{0}})$ (see \autoref{thm:tangent:decompose}), we know that the distance from $T_H$~(which is $\limexpH\ket{\Psi_{0}}$) to $\mathcal{T}(\vec{O}\oplus P',\vec{\Lambda}\oplus0,\ket{\Psi_{0}})$ is less than that to $\mathcal{T}(\vec{O},\vec{\Lambda},\ket{\Psi_{0}}$. This is contradictory to the assumption and thus completes the proof.
\end{proof}

In \autoref{thm:evolve:in:Ps}, we point out that the objective direction $\limexpH\ket{\Psi_{0}}$ must fall in the $\spacespan$ of all the new tangent spaces generated by appending a Pauli word in the Hamiltonian.
\begin{lemma}
\label{thm:evolve:in:Ps} 
\item
\begin{equation}
\limexpH\ket{\Psi_{0}}\in\spacespan\{\mathcal{T}(P',0,G(\vec{O},\vec{\Lambda})\ket{\Psi_{0}})|P'\in\mathbf{H}\}
\end{equation}
\end{lemma} 
\begin{proof}
\item
\begin{equation}
\limexpH\ket{\Psi_{0}}=-iHG(\vec{O},\vec{\Lambda})\ket{\Psi_{0}}=\sum_{P_{j}\in\mathbf{H}}-i\lambda_{j}P_{j}G(\vec{O},\vec{\Lambda})\ket{\Psi_{0}}=\sum_{P_{j}\in\mathbf{H}}\lambda_{j}\lim_{\tlim}\frac{1}{\delta t}e^{-iP_{j}\delta t}G(\vec{O},\vec{\Lambda})\ket{\Psi_{0}},
\end{equation}
which is a linear combination of vectors in $\spacespan\{\mathcal{T}(P',0,G(\vec{O},\vec{\Lambda})\ket{\Psi_{0}})|P'\in\mathbf{H}\}$. 
\end{proof}

In \autoref{thm:evolve:in:T0}, we prove that if appending single Pauli words in $\mathbf{H}$ cannot decrease the distance of the tangent space of the current product formula to the objective direction $\limexpH\ket{\Psi_0}$, then the objective direction must have been included in the tangent space. 

In other words, if $\Delta\neq 0$ (which implies the objective direction is not included in the tangent space of the current product formula), there must be a Pauli word in $\mathbf{H}$ by appending which to the product formula the distance can be decreased.

\begin{lemma} \label{thm:evolve:in:T0}
If the distance from $\limexpH\ket{\Psi_{0}}$ to $\mathcal{T}(\vec{O}\oplus P',\vec{\Lambda}\oplus0,\ket{\Psi_{0}})$
is the same as to $\mathcal{T}(\vec{O},\vec{\Lambda},\ket{\Psi_{0}})$
for all $P'\in \mathbf{H}$, then $\limexpH\ket{\Psi_{0}}$
is in $\mathcal{T}(\vec{O},\vec{\Lambda},\ket{\Psi_{0}})$. 
\end{lemma}
\begin{proof}
According to \autoref{thm:two:orthogonal}, for any $P'\in \mathbf{H}$, $T_{P'}\in\mathcal{T}(P',0,G(\vec{O},\vec{\Lambda})\ket{\Psi_{0}})$
that is orthogonal to $\mathcal{T}(\vec{O},\vec{\Lambda},\ket{\Psi_{0}})$,
$T_{P'}$ must be also orthogonal to $\limexpH\ket{\Psi_{0}}$.

Because the sum of any vectors that are orthogonal to a vector will
be still orthogonal to that vector, for any vector $T'$ in the space
$\mathcal{T}_H=\spacespan\{\mathcal{T}(P',0,G(\vec{O},\vec{\Lambda})\ket{\Psi_{0}})|P'\in\mathbf{H}\}$,
if $T'$ is orthogonal to $\mathcal{T}(\vec{O},\vec{\Lambda},\ket{\Psi_{0}})$,
$T'$ must also be orthogonal to $\limexpH\ket{\Psi_{0}}$.

Notice that we can decompose $\limexpH\ket{\Psi_{0}}$ into two parts
as 
\begin{equation}
\limexpH\ket{\Psi_{0}}=T_{\parallel}+T_{\perp},
\end{equation}
where $T_{\parallel}$ is in $\mathcal{T}(\vec{O},\vec{\Lambda},\ket{\Psi_{0}})$
and $T_{\perp}$ is orthogonal to $\mathcal{T}(\vec{O},\vec{\Lambda},\ket{\Psi_{0}})$.
However, in \autoref{thm:evolve:in:Ps} we proved that 
\[
\limexpH\ket{\Psi_{0}}\in\spacespan\{\mathcal{T}(P',0,G(\vec{O},\vec{\Lambda})\ket{\Psi_{0}})|P'\in\mathbf{H}\}=\mathcal{T}_H.
\]
Therefore, $T_{\perp}$, which is orthogonal to $\mathcal{T}(\vec{O},\vec{\Lambda},\ket{\Psi_{0}})$,
must be orthogonal to $\limexpH\ket{\Psi_{0}}$. This implies $T_{\perp}=0$
and $\limexpH\ket{\Psi_{0}}=T_{\parallel}\in\mathcal{T}(\vec{O},\vec{\Lambda},\ket{\Psi_{0}})$. 
\end{proof}
In the following, we prove that, in the ``add new operators'' procedure of adaptive PF, there always exists an operation in $\mathbf{H}$,
such that appending it to $\vec{O}$ will decrease $\Delta$ if $\Delta>0$.
\begin{proposition} \label{thm:must_decrease}
Suppose the Hamiltonian of the time evolution
is $H=\sum_{j}a_{j}P_{j}$, $a_{j}\in\mathbf{R}$. If 
\begin{equation}
\Delta'=\Delta(\vec{O},\vec{\Lambda},\ket{\Psi_{0}})=\Delta(\vec{O}\oplus P',\vec{\Lambda}\oplus0,\ket{\Psi_{0}})
\end{equation}
for all the Pauli word $P'$ in $\{P_{j}\}$, then $\Delta'=0$. 
\end{proposition} 
\begin{proof}
According to \autoref{thm:distance:error},  $\Delta(\vec{O},\vec{\Lambda},\ket{\Psi_{0}})=\Delta(\vec{O}\oplus P',\vec{\Lambda}\oplus0,\ket{\Psi_{0}})$ implies the distance from $\limexpH\ket{\Psi_{0}}$ to $\mathcal{T}(\vec{O},\vec{\Lambda},\ket{\Psi_{0}})$ is equal to that to $\mathcal{T}(\vec{O}\oplus P',\vec{\Lambda}\oplus0,\ket{\Psi_{0}})$. Because the above statement holds for all the Pauli word $P'\in \{P_i\} = \mathbf{H}$, according to \autoref{thm:evolve:in:T0}, $\limexpH\ket{\Psi_{0}}$ is in $\mathcal{T}(\vec{O},\vec{\Lambda},\ket{\Psi_{0}})$. Therefore the distance between $\limexpH\ket{\Psi_{0}}$ and $\mathcal{T}(\vec{O},\vec{\Lambda},\ket{\Psi_{0}})$ is zero and
\begin{equation}
\Delta'= \Delta(\vec{O},\vec{\Lambda},\ket{\Psi_{0}}) = 0.
\end{equation}
\end{proof}
Next, we will prove each Pauli word is only needed to appear once in an operator-adding.
\begin{proposition}\label{thm:most:add:once}
For any Pauli word $P'$, adding it twice to $\vec{\Lambda}$ will result in the same error as
\begin{equation}
\Delta(\vec{O}\oplus P'\oplus\vec{O'},\vec{\Lambda}\oplus 0\oplus\vec{0},\ket{\Psi_{0}}) = \Delta(\vec{O}\oplus P'\oplus\vec{O'}\oplus P',\vec{\Lambda}\oplus 0\oplus\vec{0}\oplus 0,\ket{\Psi_{0}}).
\end{equation}
\end{proposition}
\begin{proof}
By \autoref{thm:tangent:decompose}, we have
\begin{equation}
	\begin{split}
		& \mathcal{T}(\vec{O}\oplus P'\oplus\vec{O'},\vec{\Lambda}\oplus 0\oplus\vec{0},\ket{\Psi_{0}}) \\ 
		= & \spacespan\{\mathcal{T}(\vec{O},\vec{\Lambda},\ket{\Psi_{0}}),\mathcal{T}(P',0,G(\vec{O},\vec{\Lambda})\ket{\Psi_{0}}),\mathcal{T}(\vec{O'},\vec 0,G(\vec{O},\vec{\Lambda})\ket{\Psi_{0}})\} 
	\end{split}
\end{equation}
and
\begin{equation}
	\begin{split}
		& \mathcal{T}(\vec{O}\oplus P'\oplus\vec{O'}\oplus P',\vec{\Lambda}\oplus 0\oplus\vec{0}\oplus 0,\ket{\Psi_{0}}) \\
		= & \spacespan\{ \mathcal{T}(\vec{O},\vec{\Lambda},\ket{\Psi_{0}}),\mathcal{T}(P',0,G(\vec{O},\vec{\Lambda})\ket{\Psi_{0}}),\mathcal{T}(\vec{O'},\vec 0,G(\vec{O},\vec{\Lambda})\ket{\Psi_{0}}) , \mathcal{T}(P',0,G(\vec{O},\vec{\Lambda})\ket{\Psi_{0}}) \} \\
		= & \spacespan\{\mathcal{T}(\vec{O},\vec{\Lambda},\ket{\Psi_{0}}),\mathcal{T}(P',0,G(\vec{O},\vec{\Lambda})\ket{\Psi_{0}}),\mathcal{T}(\vec{O'},\vec 0,G(\vec{O},\vec{\Lambda})\ket{\Psi_{0}})\}  \\
		= & \mathcal{T}(\vec{O}\oplus P'\oplus\vec{O'},\vec{\Lambda}\oplus 0\oplus\vec{0},\ket{\Psi_{0}}).
	\end{split}
\end{equation}
Therefore, the distance from $\limexpH\ket{\Psi_{0}}$ to the above two spaces must be the same. This means $\Delta(\vec{O}\oplus P'\oplus\vec{O'},\vec{\Lambda}\oplus 0\oplus\vec{0},\ket{\Psi_{0}}) = \Delta(\vec{O}\oplus P'\oplus\vec{O'}\oplus P',\vec{\Lambda}\oplus 0\oplus\vec{0}\oplus 0,\ket{\Psi_{0}})$ by \autoref{thm:distance:error}.
\end{proof}
\begin{theorem} 
\label{thm:main-result}
The ``add new operators'' procedure in the jointly-optimized adaptive product formula method (Protocol 2)
satisfies the following properties. \\
 (1) The error $\Delta$ strictly decreases at each iteration until
$0$;\\
 (2) Each Pauli word is only needed to be added once;\\
 (3) An error $\Delta\leq\dcut$ can be achieved in at most $L$ iteration
for any $\dcut\geq0$.  \end{theorem}
\begin{proof}
According to \autoref{thm:must_decrease}, as long as $\Delta\neq 0$,
there exists a $P'\in\mathbf{H}$ that can decrease $\Delta$ after being
appended to the product formula. Thus, we prove (1). 

By \autoref{thm:most:add:once}
(where we prove (2)), in an operator-adding procedure, every $P'\in\mathbf{H}$ can
only be appended and decrease $\Delta$ of the product formula once. Therefore,
the operator-adding will stop when all the $P'$ in $\mathbf{H}$ are appended to the product formula. Thus,
the operator-adding will have at most $L$ iterations, as there are $L$ Pauli words in $\mathbf{H}$. Also, if the
operator-adding ends after $L$ iterations, the $\Delta$ of
the final product formula must be $0$ according to \autoref{thm:must_decrease}
as no Pauli word in $\mathbf{H}$ will decrease $\Delta$.
Therefore, $\Delta\leq\dcut$ can be achieved in the ``add new operators'' procedure by
at most $L$ iterations for any $\dcut\geq 0$. Thus, we prove (3). 
\end{proof}

\section{Shot noise analysis}
\label{SM:shotnoiseanalysis}
\newcommand{\eres}{\epsilon_{\mathrm{res}}}
\newcommand{\tocheck}{{\color{red}(check)}}
\newcommand{\vlam}{\vec{\lambda}}

In practice, all the expectation values that we measure with quantum circuits are estimated with a finite number of measurements. In this section, we analyze the effect of statistical noise, or called shot noise, on the performance of our method.
We propose a measurement strategy given finite shots, and we discuss the number of shots that are required in each step of our method. 

\textbf{Overview:} In \autoref{SM:noise-effect-on-delta}, we give the basic setup of our analysis and give a relation between the variance of the estimated $\Delta^2$ and the variance of the estimated $A, C$. In \autoref{SM:measure-AC}, we show explicitly how the terms of $A$ and $C$ can be measured on a quantum computer given a certain number of shots and we show the measurement complexity of $C$ can be independent of the number of terms in the Hamiltonian.
In \autoref{SM:optimal-shot-allocation}, we show how to optimize the measurement distribution, based on which we can estimate the required number of measurements given a required bound of the error of $\Delta^2$ or $\Delta$. 
Finally, in \autoref{SM:scalability}, we propose a method to suppress $\|\vec{\lambda}\|_1$ in our method while maintaining the accuracy, since a large $\|\vec{\lambda}\|_1$ will induce a large measurement cost.
We also provide analysis of measurement complexity and simulation accuracy in \autoref{SM:scalability}.

\subsection{Effect of shot noise on $\Delta'^2$}
\label{SM:noise-effect-on-delta}
We first state our assumptions used in the analysis. In Protocol 2, three quantities are directly measured with quantum circuits:  matrix $A$, vector $C$, and $\langle H^2 \rangle$. 
Note that $\langle H^2 \rangle$ is a value that should be invariant under an ideal time evolution $U = e^{-iHt}$, and thus we may not need to measure it frequently given limited measurement resources.  We note that $\langle H^2 \rangle$ may also be classically calculated if the initial state has a compact classical representation (e.g., Hartree-Fock state).
Below, $\langle H^2 \rangle$ is assumed to be a constant, and we only consider the statistical noise on $A$ and $C$. 

Denote the random variables of the estimation of $A$ and $C$ by $A'$ and $C'$. As it is assumed that there are only statistical errors and the entries of $A'$ and $C'$ are directly measured on quantum circuits, the entries of $A'$ and $C'$ are unbiased, i.e., their expectation values are true values.
We also note that the entries of $A'$ and $C'$ are independent of each other because the measurement processes for them are independent, except for the symmetric entries in $A$ which are equal ($A_{ij}=A_{ji}$).

Recall from \autoref{equ:SM_Delta} that in the ``joint parameter optimization'' procedure, the error $\Delta^2$ with a certain $\vec{\lambda}'$ in the product formula can be written as
\begin{equation}
    \Delta^2(\vec{\lambda}') = \braket{H^2} + \vec{\lambda}'^{T}A \vec{\lambda}' -2\vec{\lambda}'^{T} C.
\end{equation}
In the noiseless case, as true values $A$ and $C$ are assumed to be available by measurements, one can directly obtain $\vlam'$ that minimizes $\Delta^2(\vec{\lambda}')$ by solving $A\vlam'=C$. 
In the presence of noise, 
as $A'$ and $C'$ deviate from the true value,
if one naively adopts the $\vec{\lambda}'$ that is solved from the equation $A'\vec{\lambda}'=C'$, it is possible that $\vec{\lambda}'$ deviates largely from the solution of $A\vec{\lambda}=C$ even when the noise on $A'$ is small. 
The problem is especially severe when 
$A$ is ill-conditioned.
From another point of view, in the presence of noise, the estimation of $\Delta^2(\vec{\lambda}')$ for each $\vlam'$ deviates from its true value, and hence we get a noisy objective $\Delta'^2$, which is different from $\Delta^2$. By solving $A'\vec{\lambda}'=C'$, one adopts $\vlam'$ that minimizes the noise-affected objective, which might deviate largely from the true objective $\Delta^2$ at some $\vlam'$.
We note that adopting a bad $\vec{\lambda}'$, which gives a large $\Delta^2(\vec{\lambda}')$, will induce a large simulate error. 
Therefore, instead of directly solving $A'\vec{\lambda}'=C'$, one needs a more effective strategy to infer $\vec{\lambda}'$ that can minimize $\Delta^2(\vec{\lambda}')$, provided the observed $A'$ and $C'$. 

In this section, we take account of the uncertainty in an estimation of $\Delta^2(\vec{\lambda}')$. We will first discuss how to estimate this uncertainty and then discuss how to obtain $\vlam'$ such that the uncertainty in the estimation of $\Delta^2(\vec{\lambda}')$ is suppressed and meanwhile the objective $\Delta^2(\vlam')$ is nearly minimized.
Let us consider the following random variable which describes the estimation of $\Delta^2(\vec{\lambda}')$ with noisy values $A'$ and $C'$,
\begin{equation}
  \Delta'^2(\vec{\lambda}') = \braket{H^2} + \vec{\lambda}'^{T}A'  \vec{\lambda}' -2\vec{\lambda}'^{T} C'.
\end{equation}
Every time   $A$ and $C$ are measured and  $\Delta^2(\vec{\lambda}')$ is calculated by the observed values of $A'$ and $C'$, one obtains an estimation of  $\Delta^2(\vec{\lambda}')$, which is {a sample that obeys the distribution of the random variable $\Delta'^2(\vec{\lambda}')$.} We can estimate the uncertainty in an estimation of $\Delta^2(\vlam')$  by the distribution of $\Delta'^2(\vec{\lambda}')$.
If the distribution has a small variance and is concentrated, as $A'$ and $C'$ are unbiased, a sample of $\Delta'^2(\vec{\lambda}')$ will be close to the true value. In this case, if an instance, sampled from the distribution of $\Delta'^2(\vec{\lambda}')$, is less than $\dcut-\epsilon$ with $\epsilon$ being a value related to the variance, we can use $\vlam'$ to update the parameters in the product formula.
In contrast, if the variance of $\Delta'^2(\vec{\lambda}')$ is large, there is a high probability that a sample of $\Delta'^2(\vec{\lambda}')$ deviates from the true value $\Delta^2(\vec{\lambda}')$. In this case, adopting $\vlam'$  to update the parameters is risky because $\Delta^2(\vec{\lambda}')$ may be above the cutoff $\dcut$. 

To illustrate this intuition, we give an explicit form of the variance of $\Delta'^2(\vec{\lambda}')$ provided the variances of $A'$ and $C'$, which can be obtained from the number of measurements in practice.
Notice that $\Delta'^2(\vec\lambda')$ is a summation of independent random variables, and thus the variance of $\Delta'^2(\vec{\lambda}')$ is given by
\begin{align}
\label{equ:variance-Delta2}
    \var [\Delta'^2(\vec{\lambda}')] =\sum_{i,j>i} (2 \lambda'_i \lambda'_j)^2 \var [A'_{ij}] + \sum_{j}  (2\lambda'_j)^2  \var [C'_{j}].
\end{align}
By Chebyshev's inequality, we have
\begin{align}
    \Pr\bigg(|\Delta'^2(\vec{\lambda}')-\Delta^2(\vec{\lambda}')|\geq \sqrt{ \frac{\var[\Delta'^2(\vec{\lambda}')]}{\delta}}\bigg) \leq \delta,
\end{align}
from which one can estimate the error of each estimation of $\Delta^2(\vec{\lambda}')$.

Next, we show how to obtain $\vlam'$ such that the true value $\Delta^2(\vlam')$ and the variance of $\Delta'^2(\vlam')$ are both small. To address this, we first give a pessimistic bound for $\var[\Delta'^2(\vlam')]$,
\begin{align}
\label{equ:pessi-bound-first-appear}
\var[\Delta'^2(\vlam')]
< \frac{1}{n_M}\Big(\|\vec{\lambda}'\|_1^2+2\|a\|_1\|\vec{\lambda}'\|_1\Big)^2.
\end{align}
The derivation of the bound can be found in \autoref{SM:circuit-scalability}. The above bound shows that one can reduce the variance of $\Delta'^2(\vlam')$ by reducing the norm $\|\vec{\lambda}'\|_1$. Therefore, a natural idea is to modify the 
objective $\Delta^2(\vlam')$ so that $\vlam'$ with a smaller $l_1$ norm can be obtained. Suppose $A^*$ and $C^*$ are two samples obtained from the distributions of $A'$ and $C'$ in an experiment. We consider a modified objective
\begin{equation}
\label{equ:regularized-Delta-obj}
  \Delta^{*2}_{\mathrm{reg}}(\vec{\lambda}') = \braket{H^2} + \vec{\lambda}'^{T}A^*  \vec{\lambda}' -2\vec{\lambda}'^{T} C^* +\beta\|\vec{\lambda}'\|_1,
\end{equation}
where $\beta$ is an adjustable positive number. By including the $l_1$ regularization term $\beta\|\vec{\lambda}'\|_1$, $\vlam'$ that minimizes the above objective tends to have a small $l_1$ norm and therefore the number of measurements to achieve a certain value of $\var[\Delta'^2(\vlam')]$ will be decreased according to \autoref{equ:pessi-bound-first-appear}.
With $\beta$ and $\var[\Delta'^2(\vlam')]$ being small, it can be expected that $\vlam'$ minimizes $\Delta^{*2}_{\mathrm{reg}}(\vec{\lambda}')$ also nearly minimizes $\Delta^{2}(\vec{\lambda}')$.

We note that $\Delta^{*2}_{\mathrm{reg}}(\vec{\lambda}')$ can be minimized in a similar way to $\Delta^2(\vlam')$. Consider the $\vec{\lambda}'$ that makes the derivative of $\Delta^{*2}_{\mathrm{reg}}(\vec{\lambda}')$ concerning $\vec{\lambda}'$ equals $0$, that is,
\begin{equation}
    \frac{\partial \Delta^{*2}_{\mathrm{reg}}(\vec{\lambda}')}{\partial \lambda'_j} = 2\sum_{i} A^{*}_{ij}\lambda'_i - 2C^{*}_j+\beta \lambda'_j/|\lambda'_j|=0.
\end{equation}
$\vlam'$ that minimizes $\Delta^{*2}_{\mathrm{reg}}(\vec{\lambda}')$ satisfies the following equation,
\begin{equation}
    \sum_{i} A^{*}_{ij}\lambda'_i = C^{*}_j-\frac{\beta}{2} \lambda'_j/|\lambda'_j|.
\end{equation}
The above equation can be solved with a correct initial guess of the signs (i.e., $\lambda'_j/|\lambda'_j|$) of $\vlam'$. Otherwise, several iterations might be required in order to get a self-consistent solution.

In what follows, we first show how to estimate $\var [A'_{ij}]$ and $\var[C_{j}']$ in experiments, so that $\var [\Delta'^2(\vec{\lambda}')]$ can be estimated. Then, we discuss how to optimize the measurement cost with an efficient shot allocation strategy. After that, as $\|\vlam'\|_1$ is closely related to the variance, we demonstrate how to suppress $\|\vlam'\|_1$ and hence reduce the measurement cost. Based on the improved protocol, we analyze the scalability of our method. Finally, we discuss how we can estimate the error of $\Delta'(\vlam')$ from the variance of $\Delta'^2(\vlam')$.

\subsection{Strategies for measurements}
\label{SM:shot-optimize}

In the above discussion, we showed how $\Delta'^2(\vlam')$ is related to the variance of the entries of $A'$ and $C'$.
In the following, we first derive bounds of $\var[A'_{ij}]$ and $\var[C_{j}']$ given a certain number of shots. 
Then, we show the relation between the variance, $\var [\Delta'^2(\vlam')]$, and the number of shots with an optimized shot allocation strategy.

\subsubsection{Measurement of $A$ and $C$}
\label{SM:measure-AC}
We first illustrate how the entries of $A$ and $C$ can be measured on a quantum circuit.  $A_{k_1k_2}$ ($k_1<k_2$) can be measured by the circuit below \cite{li2017efficient,yuan2019theory}.
\begin{align}
\label{equ:hamil-innerp-A}
\Qcircuit @C=1.5em @R=2.0em {
&\lstick{\ket{0}}&\gate{H}&\qw  & \ctrlo{1}    & \qw&  \qw&  \ctrl{1}&\measure{\mbox{\;X\;}} \\
&\lstick{\ket{\Psi_0}}& \qw  & \gate{\prod_{j=1}^{k_1} e^{-iP_j\Lambda_j}}&\gate{-iP_{k_1}} &\gate{\prod_{j=k_1+1}^{k_2} e^{-iP_j\Lambda_j}} &   \qw& \gate{-iP_{k_2}} &\qw\\
}
\end{align}
The variance of the one-shot estimation is ($\mathbb{E}$ denotes the expectation value)
\begin{equation}
    \var[A'_{k_1k_2}]_{\mathrm{(one\;shot)}} = \mathbb{E}[M(X)^2] - \mathbb{E}[M(X)]^2 = \mathbb{E}[1] - \langle X \rangle^2 = 1 - A_{k_1k_2}^2,
\end{equation}
where  $M(X)$ denotes a one-shot outcome on the ancilla qubit. 
The variance of an average over $n_{Ak_1k_2}$ shots is
\begin{equation}
\var[A'_{k_1k_2}] = \frac{1 - A_{k_1k_2}^2}{n_{Ak_1k_2}}.
\end{equation}
The measurement of $C$ uses a similar circuit to $A$, whose output evaluates $\operatorname{Im} \left( \braket{\Psi_0 |\partial_{k} G^{\dagger}(\vec O,\vec\Lambda ) P_l  G(\vec O,\vec\Lambda)  |\Psi_0} \right)$.

\begin{align}
\label{equ:hamil-innerp-C}
\Qcircuit @C=1.5em @R=2.0em {
&\lstick{\ket{0}}&\gate{H}&  \gate{S} & \ctrl{1}    & \qw&  \qw&  \ctrlo{1}&\measure{\mbox{\;X\;}} \\
&\lstick{\ket{\Psi_0}}& \qw & \gate{\prod_{j=1}^{k} e^{-iP_j\Lambda_j}} &\gate{-iP_k} &\gate{\prod_{j=k+1}^{L} e^{-iP_j\Lambda_j}} &   \qw& \gate{P_l} &\qw\\
}
\end{align}

$C_k$ for $H=\sum_{l=1}^{L} a_l P_l$ can be evaluated by adding up the outcomes as 
\begin{align}
C_k & = \operatorname{Im} \left( \braket{\Psi_0 |\partial_{k} G^{\dagger}(\vec O,\vec\Lambda ) H  G(\vec O,\vec\Lambda)  |\Psi_0} \right)\\
&= \sum_{l=1}^{L} a_l \operatorname{Im} \left( \braket{\Psi_0 |\partial_{k} G^{\dagger}(\vec O,\vec\Lambda ) P_l  G(\vec O,\vec\Lambda)  |\Psi_0} \right).
\end{align}

The above method measures each term in the Hamiltonian one-by-one, and hence the measurement complexity scales as $\mathcal{O}(L)$.
It is not that straightforward whether advanced measurement techniques, such as classical shadows~\cite{huang2020predicting,hadfield2020measurements} and Pauli word grouping~\cite{wu2021overlapped}, can be applied in the measurement of $C$ to improve the measurement efficiency since we measure on the ancillary qubit.
In the following, we address that all these techniques for measuring $ \langle H \rangle $ can be applied in this context.

Notice that the last gate in \autoref{equ:hamil-innerp-C}, which is a controlled Pauli gate, is a Clifford gate. This means it can be absorbed into the observable without increasing its number of Pauli words. Denote the unitary of the last gate by $\mathrm{Ctrl}(P_l)$ and the state before the last gate by $\ket{\phi_k}$. With the ancilla qubit labeled by $0$,  the expectation value can be expressed as
\begin{equation}
\bra{\phi_k}\mathrm{Ctrl}(P_l)^{\dagger} X_0 \mathrm{Ctrl}(P_l) \ket{\phi_k}.
\end{equation}
$\mathrm{Ctrl}(P_l)$ can be written as 
\begin{equation}
    \mathrm{Ctrl}(P_l) = \ket{1}_0\bra{1}_0 \otimes P_l + \ket{0}_0\bra{0}_0 \otimes I,
\end{equation}
and we have
\begin{align}
\mathrm{Ctrl}(P_l)^{\dagger} X_0 \mathrm{Ctrl}(P_l) &= (\ket{1}_0\bra{1}_0 \otimes P_l + \ket{0}_0\bra{0}_0 \otimes I)(\ket{1}_0\bra{0}_0+\ket{0}_0\bra{1}_0)(\ket{1}_0\bra{1}_0 \otimes P_l + \ket{0}_0\bra{0}_0 \otimes I) \\
&=  (\ket{1}_0\bra{1}_0 \otimes P_l + \ket{0}_0\bra{0}_0 \otimes I)(\ket{0}_0\bra{1}_0 \otimes P_l + \ket{1}_0\bra{0}_0 \otimes I) \\
&=  (\ket{0}_0\bra{1}_0+\ket{1}_0\bra{0}_0) \otimes P_l = X_0 \otimes P_l
\end{align}
Therefore, we can define an operator 
\begin{equation}
H'=\sum_l a_l X_0\otimes P_l 
\end{equation}
and have
\begin{align}
    C_k &= \sum_l a_l \bra{\phi_k} \mathrm{Ctrl}(P_l)^{\dagger} X_0  \mathrm{Ctrl}(P_l) \ket{\phi_k} \\
    &= \sum_l a_l \bra{\phi_k} X_0\otimes P_l \ket{\phi_k} \\
    &= \bra{\phi_k} H' \ket{\phi_k}.
\end{align}

By the above reduction,  the problem of measuring $C_k$ is reduced to the problem of measuring $\langle H' \rangle$ on the state $\ket{\phi_k}$. As such, all the techniques for measuring an observable can be applied here. Our result directly applies to other 
quantum algorithms, in which terms similar to $C_k$ needs to be evaluated.

Finally, we show an explicit form of the variance of $C'_k$ by the $l_1$-sampling measurement strategy. The key idea is that we sample the to-be-measured Pauli word  in $H'$ according to the absolute value of their coefficients. With $P_j$ being a Pauli word to be sampled and $M(P_j)$ being the outcome of the one-shot measurement ($1$ or $-1$), the one-shot estimation of the Hamiltonian is given by
\begin{equation}
    \|a\|_1\mathrm{sign}(a_j)M(P_j),
\end{equation}
whose expectation value is unbiased
\begin{align}
    &\mathbb{E}_l[\mathbb{E}_{M(P_l)}[ \|a\|_1\mathrm{sign}(a_l)M(P_l)]]
    = \sum_l\frac{|a_l|}{\|a\|_1} \|a\|_1\mathrm{sign}(a_j)\langle P_l \rangle
    = \sum_l a_l \langle P_l \rangle = \langle H' \rangle.
\end{align}
The variance of a one-shot estimation is
\begin{align}
\var[C'_k]_{\mathrm{(one\;shot)}} &= \mathbb{E}[(\|a\|_1\mathrm{sign}(a_j)M(P_j))^2]-\mathbb{E}[\|a\|_1\mathrm{sign}(a_j)M(P_j)]^2 \\
&= \|a\|_1^2- \langle H' \rangle^2 =  \|a\|_1^2- C_k^2,
\end{align}
and the variance of an average over $n_{Ck}$ shots naturally follows
\begin{align}
\var[C'_k] = \frac{\|a\|_1^2- C_k^2}{n_{Ck}}.
\end{align}

\subsubsection{Optimal Shot allocation}
\label{SM:optimal-shot-allocation}

In this section, we discuss an efficient shot allocation strategy for measuring the entries of $A$ and $C$.
With the variance we deduced in \autoref{SM:measure-AC}, we can rewrite \autoref{equ:variance-Delta2} as
\begin{align}
\label{equ:naive-variance}
\var[\Delta'^2(\vlam')]
=\sum_{i,j>i} 4(\lambda'_{i}\lambda'_{j})^2\frac{1-A_{ij}^2}{n_{Aij}} + \sum_{j} 4(\lambda'_{j})^{2}\frac{\|a\|_1^2-C_j^2}{n_{Cj}}.
\end{align}
The variance is minimized when 
\begin{equation}
\label{equ:shot-allocation}
n_{Aij}  = M \sqrt{4(\lambda'_{i}\lambda'_{j})^2(1-A_{ij}^2)} ~~,~~ n_{Cj} = M \sqrt{4(\lambda'_{j})^{2}(\|a\|_1^2-C_j^2)},  
\end{equation}
where $M$ is a normalization factor which can be determined by $\sum_{i,j>i}n_{Aij} + \sum_j n_{Cj} = n_M$ and $n_M$ is the total number of shots. With this allocation, we have
\begin{align}
\label{equ:var-A-C-lam}
\var[\Delta'^2(\vlam')] & = \frac{1}{n_M}\bigg(\sum_{i,j>i}  \sqrt{4(\lambda'_{i}\lambda'_{j})^2(1-A_{ij}^2)} + \sum_j \sqrt{4(\lambda'_{j})^{2}(\|a\|_1^2-C_j^2)}\bigg)^2\\
&=\frac{4}{n_M}\bigg(\sum_{i,j>i} |\lambda'_{i}\lambda'_{j}| \sqrt{(1-A_{ij}^2)} + \sum_j |\lambda'_{j}|\|a\|_1\sqrt{(1-C_j^2/\|a\|_1^2)}\bigg)^2.
\end{align}

It is difficult to allocate the shots optimally in practice because   $A$, $C$ and the associated $\vlam'$ are unknown in advance. In the case where $A$, $C$ are unknown and one wants to minimize the variance with a given $\vec{\lambda}'$. A strategy is that we can allocate the shots as if the entries of $A$ and $C$ are all zero in \autoref{equ:shot-allocation}, and the variance can be bounded by
\begin{equation}
\label{equ:var-delta-bound-zero-A-C}
\var[\Delta'^2(\vlam')] \leq \frac{4}{n_M}(\sum_{i,j>i} |\lambda'_{i}\lambda'_{j}| + \sum_j |\lambda'_{j}|\|a\|_1)^2.
\end{equation}

In practice, a more common case is where $A$, $C$ and the concerned $\vec{\lambda}'$ are all unknown before the shot allocation. We show that in this case, the shot allocation can be done with a guess of $\|\vlam\|_1$. In \autoref{SM:circuit-scalability}, we will show that, with a fixed $\|\vec{\lambda}'\|_1$, the bound in \autoref{equ:var-delta-bound-zero-A-C} takes its maximum when the entries of $\vec{\lambda}'$ are all the same. This gives an upper bound
\begin{align}
\label{equ:var-delta-bound-uniform-lambda}
\var[\Delta'^2(\vlam')]
< \frac{1}{n_M}(\|\vec{\lambda}'\|_1^2+2\|a\|_1\|\vec{\lambda}'\|_1)^2,
\end{align}
which only depends on $\|\vec{\lambda}'\|_1$ and $\|a\|_1$. Therefore, one can distribute the measurements by assuming that $\lambda'_j=\|\vec{\lambda}'\|_1/n_O$ in \autoref{equ:shot-allocation} ($n_O$ is the dimension of $\vec{\lambda}'$). In \autoref{SM:reduce-lambda-norm}, we will show one can always find an acceptable solution $\vec{\lambda}'$ whose $l_1$ norm is less than or equal to $\|a\|_1$ by adding operators to the product formula.

\subsubsection{Discussions}

A non-zero $\var[\Delta'^2(\vlam')]$ results in a deviation of $\Delta'^2(\vlam')$ from the true value $\Delta^2(\vlam')$. One can use Chebyshev's inequality or Hoeffding's inequality to derive the interval of $\Delta'^2(\vlam')$ given a failure probability $\delta$ (See \autoref{SM:failure-probability}). Then, the interval of $\Delta'^2(\vlam')$ can be used to derive the interval for $\Delta'(\vlam')$. 
The upper bound of $\Delta'(\vlam')$ can be used to calculate $\Delta^{(\max)}$ and estimate the simulation error. Based on this, one can estimate the required number of measurements to achieve a certain simulation accuracy.

Another effect of the variance of $\Delta'^2(\vlam')$ is that in the ``add new operators'' procedure where we use $\Delta'(\vlam')$ to decide which operator to add to the circuit, an inaccurate $\Delta$ makes the choice less optimal and make the generated circuit deeper.

\subsection{Scalability}
\label{SM:scalability}
In this section, we analyze the scalability of our method.
As we discussed in the previous sections, 
the magnitudes of the entries of $\vec{\lambda}'$ are
important for estimating the variance. Therefore, we first address that by adaptively adding new operators, there exists a solution $\vec{\lambda}'$, which satisfies  $\|\vec{\lambda}' \|_1 \leq \|a\|_1$ and $\Delta(\vlam')\leq \dcut/2$. Then, we discuss the measurement complexity and the error scaling  with a bounded $\|\vec{\lambda}'\|_1$.

\subsubsection{Reduce $\|\vec{\lambda}\|$ by adding more operators}
\newcommand{\lamOldvec}{\vec{\lambda}^{\mathrm{old}}}
\newcommand{\lamNewvec}{\vec{\lambda}^{\mathrm{new}}}

\newcommand{\lamOld}{\lambda^{\mathrm{old}}}
\newcommand{\lamNew}{\lambda^{\mathrm{new}}}

\label{SM:reduce-lambda-norm}
In \autoref{equ:var-A-C-lam}, we showed that large entries in a $\vec{\lambda}'$ will induce a large variance. 
However, when the circuit construction is improper, it is possible that all the $\vec{\lambda}'$, which give a small $\Delta^2$, have large entries, so that the measurement cost of the method is large. In fact, this is a severe problem for previous methods that use a fixed circuit, because it is difficult to know whether such a bad set of $A, C$ will appear for a given parameterized circuit and initial state.

Fortunately, the problem of large entries can be mitigated by our adaptive circuit construction strategy. In the following, we consider the noiseless case and show that by adaptively adding new operators to the product formula, one can always obtain a numerically good circuit, characterized by a good $\vec{\lambda}'$, satisfying $\Delta(\vec{\lambda}')\leq \dcut/2$ and $\|\vec{\lambda}'\|_1 \leq \|a\|_1$ ($\|a\|_1$ is the weight of the Hamiltonian). 
Finally, at the end of the section, we discuss the implication of the result in the presence of noise.

In Protocol 2, the operator-adding procedure terminates when $\vlam'$ with $\Delta(\vlam')\leq \dcut/2$ is obtained. However, such $\vlam'$ might have a large $l_1$ norm. 
To solve this problem, as we do not require $\Delta(\vlam')$ being further lower than $\dcut/2$, we focus on a set of $\vlam'$ whose $\Delta(\vlam')$ is less than $\dcut/2$ and study their norm.
The following observation implies that adding a new operator to a product formula does not increase the minimum $l_1$ norm among the set of $\vlam'$ which gives $\Delta(\vlam')\leq \dcut/2$.

\begin{observation}
\label{thm:lambda-norm-can-decrease}
In an ``add new operators'' procedure in Protocol 2, suppose there is a solution $\lamOldvec$. After adding a new operator to the current product formula, there must exist a $\lamNewvec_{}$, such that $\Delta_{\mathrm{new}}(\lamNewvec_{})=\Delta_{\mathrm{old}}(\lamOldvec)$ and $\|\lamNewvec_{}\|_1\leq \|\lamOldvec\|_1$.  Here, $\Delta_{\mathrm{old}}(\cdot)$ and $\Delta_{\mathrm{new}}(\cdot)$ denote the function of $\Delta$ concerning $\vec{\lambda}$ in the old and new product formula.
\end{observation}

\begin{proof}
Given $\lamOldvec$, the observation can be easily proven by setting $\lamNewvec_{}=\lamOldvec\oplus 0$. Because of $\|\lamOldvec\oplus 0\|_1 = \|\lamOldvec\|_1$, it is obvious that $\|\lamOldvec\oplus 0\|_1 \leq \|\lamOldvec\|_1$. To verify $\Delta_{\mathrm{new}}(\lamOldvec\oplus 0)=\Delta_{\mathrm{old}}(\lamOldvec)$, one can simply expand $\Delta_{\mathrm{new}}(\lamOldvec\oplus 0)$ and $\Delta_{\mathrm{old}}(\lamOldvec)$ by \autoref{equ:SM_Delta}. Alternatively, one just needs to recall \autoref{def:tangent:pf} and it is easy to find that $\lamOldvec$ and $\lamOldvec \oplus 0$ correspond to the same vector in the tangent space. Thus, the distances between their corresponding vectors to the vector for the ideal evolution are the same. This implies their $\Delta$ are the same.
\end{proof}

The above observation implies that the minimal norm among all the acceptable $\vec{\lambda}'$ whose $\Delta$ is smaller than $\dcut/2$ will not be increased by adding operators. 
However, the minimal norm could remain the same. In the following, we show that we can suppress the minimal $\| \vec{\lambda}'\|_1$ by adding new Pauli words when it is larger than $\|a\|_1$. 

\begin{observation}
\label{thm:decrease-to-a-norm}
In an ``add new operators'' procedure in Protocol 2, if all the Pauli words in the Hamiltonian have been added to the product formula in the current ``add new operators'' procedure, there exists a solution $\vec{\lambda}'$ such that $\|\vec{\lambda}'\|_1 \leq \|a\|_1$ and $\Delta(\vec{\lambda}')=0$ (therefore $\Delta(\vec{\lambda}')\leq \dcut/2$). Here, $\{a_j\}$ are the coefficients of the Pauli words in the Hamiltonian. 
\end{observation}

\begin{proof}
This observation can be verified by proving that: when $\vec{\lambda}' = (0,0,\cdots,0,a_1, a_2, \cdots, a_L)$ and $\vec{O}=(\cdots,P_1, P_2, \cdots, P_L)$, $\Delta(\vlam')=0$. This completes the proof because of $\|\vec{\lambda}'\|_1=\|a\|_1$. We first show that $A\vec{\lambda}'=C$.  Notice that $\lambda'_{i}=0$ for $i\leq n_O-L$ and $\lambda'_{n_O-L+j}=a_j$. From the definition in \autoref{equ:A-C-define},  the $k$th entry of $A\vec{\lambda}'$ is given by
\begin{align}
    (A\vec{\lambda}')_k&=\sum_{j=1}^{n_O} A_{jk}\lambda'_{j}=\sum_{j=n_O-L+1}^{n_O} A_{jk}\lambda'_{j}
    = \sum_{j=n_O-L+1}^{n_O}  \operatorname{Re}\left( \bra{\Psi_0}\partial_{k}G^{\dagger}(\vec O,\vec\Lambda )  \partial_{j}G(\vec O,\vec\Lambda  )\ket{\Psi_0} \right) \lambda'_{j}\\
    &=   \sum_{j=1}^{L}a_{j} \operatorname{Re}\left( \bra{\Psi_0}\partial_{k}G^{\dagger}(\vec O,\vec\Lambda )   (-iP_j) G(\vec O,\vec\Lambda  )\ket{\Psi_0} \right) 
    =  \operatorname{Re}\Big( -i \bra{\Psi_0}\partial_{k}G^{\dagger}(\vec O,\vec\Lambda )  \sum_{j=1}^{L}(a_{j}P_j) G(\vec O,\vec\Lambda  )\ket{\Psi_0} \Big) \\
    &=\operatorname{Im} \Big( \braket{\Psi_0 |\partial_{k} G^{\dagger}(\vec O,\vec\Lambda ) H  G(\vec O,\vec\Lambda)  |\Psi_0} \Big) = C_k,
\end{align}
which proves $A\vec{\lambda}'=C$. Therefore, we have $\Delta^2=\expectv{H^2}-\vec{\lambda}'^TC.$ By the same trick, we have
\begin{align}
    \vec{\lambda}'^TC &= \sum_{j=1}^{L} a_j \operatorname{Im} \left( \braket{\Psi_0 |\partial_{n_O-L+j} G^{\dagger}(\vec O,\vec\Lambda ) H  G(\vec O,\vec\Lambda)  |\Psi_0} \right) 
    =   \sum_{j=1}^{L} \operatorname{Im}  \left( \braket{\Psi_0 |G^{\dagger}(\vec O,\vec\Lambda ) ia_jP_j H G(\vec O,\vec\Lambda)  |\Psi_0} \right)\\
    & = \operatorname{Im}  \left( i\braket{\Psi_0 |G^{\dagger}(\vec O,\vec\Lambda ) H^2 G(\vec O,\vec\Lambda)  |\Psi_0} \right) = \expectv{H^2},
\end{align}
which proves $\Delta^2=0$ and $\Delta=0$.
\end{proof}


The last observation shows that the minimal $l_1$ norm of $\Vec{\lambda}'$ with $\Delta(\Vec{\lambda}')\leq\dcut/2$ can not be reduced by adding an operator twice.

\begin{observation}
\label{thm:no-twice-for-dec-norm}
Suppose it is in an ``add new operators'' procedure in Protocol 2 and suppose Pauli word $P'$ from the Hamiltonian has been added to the product formula once in the current ``add new operators'' procedure. If $P'$ is added to the product formula again,  $\min \|\lamOldvec\|_1 \leq \min \|\lamNewvec\|_1$ will hold under the constraint that $\Delta_{\mathrm{old}}(\lamOldvec) = \Delta_{\mathrm{new}}(\lamNewvec)$. Here, $\mathrm{old}$ and $\mathrm{new}$ labels the configurations before and after $P'$ is added again.
\end{observation}
\begin{proof}
This observation can be verified by noticing that there is a transformation that keeps the same $\Delta$ and maps every $\lamNewvec$ to $\lamOldvec$ without increasing the $l_1$ norm. Suppose there are $n_O$ entries in $\lamNewvec$ and the two indices correspond to $P'$ are $p$ and $n_O$. We can define a map $T$ that maps $n_O$ dimension vectors to $n_O-1$ dimension ones, so that $T(\lamNewvec)_j$ equals to $\lamNew_{j}$ except for $j=p$ and $T(\lamNewvec)_p=\lamNew_{p}+\lamNew_{n_O}$. Because $|\lamNew_{p}+\lamNew_{n_O}| \leq |\lamNew_{p}|+|\lamNew_{n_O}|$, we can see that $\|T(\lamNewvec)\|_1 \leq \|\lamNewvec\|_1$. Therefore, the map will only decrease or keep the $l_1$ norm of the input. Also, we can prove that $\Delta_{\mathrm{old}}(T(\lamNewvec)) = \Delta_{\mathrm{new}}(\lamNewvec)$. Notice that for $i,j\in\{1,2,\dots,n_O-1\}$, we have $A^{\mathrm{old}}_{ij}=A^{\mathrm{new}}_{ij}$ and $C^{\mathrm{old}}_i=C^{\mathrm{new}}_i$. Also, notice that $A^{\mathrm{new}}_{p,n_O}=1$, $A^{\mathrm{new}}_{pj}=A^{\mathrm{new}}_{n_O, j}$ and $C^{\mathrm{new}}_{p}=C^{\mathrm{new}}_{n_O}$. 
Therefore, one can verify that
\begin{align}
    \sum_{j=1}^{n_O-1} C^{\mathrm{old}}_{j} T(\lamNewvec)_j&=\sum_{j=1}^{n_O} C^{\mathrm{new}}_{j} \lamNew_{j} ~~\mathrm{and}~~
    \sum_{i,j=1}^{n_O-1} A^{\mathrm{old}}_{ij} T(\lamNewvec)_iT(\lamNewvec)_j =\sum_{i,j=1}^{n_O} A^{\mathrm{new}}_{ij} \lamNew_{i}\lamNew_{j}.
\end{align}
Then, it can be verified with the definition in \autoref{equ:SM_Delta} that
\begin{align}
\Delta_{\mathrm{new}}^2(\lamNewvec) - \Delta_{\mathrm{old}}^2(T(\lamNewvec)) = 0.
\end{align}
Thus, we can see $T$ will not change the $\Delta$ of the input. Therefore, with any $\Delta^*>0$, for every $\vec{\lambda}^*$ such that $\|\vec{\lambda}^*\|_1 = \min_{\Delta_{\mathrm{new}}(\lamNewvec)=\Delta^*} \|\lamNewvec\|_1$, there is $\|T(\vec{\lambda}^*)\|_1 \leq \|\vec{\lambda}^*\|_1$ with $\Delta_{\mathrm{old}}(T(\vec{\lambda}^{*}))=\Delta^*$. This directly implies 
\begin{equation}
    \min_{\Delta_{\mathrm{old}}(\lamOldvec)=\Delta^*} \|\lamOldvec\|_1 \leq \min_{\Delta_{\mathrm{new}}(\lamNewvec)=\Delta^*} \|\lamNewvec\|_1.
\end{equation}
\end{proof}

\autoref{thm:lambda-norm-can-decrease} indicates that by adding a new operator, the minimal $l_1$ norm of the $\vec{\lambda}'$ that gives $\Delta(\vec{\lambda}')\leq \dcut/2$ will not increase, or equivalently, adding new operators might decrease the norm. Then, in \autoref{thm:no-twice-for-dec-norm}, we show that one does not need to add a same operator twice to decrease the norm, so the norm reduction will terminate when all the operators in the Hamiltonian are added to the product formula.   \autoref{thm:decrease-to-a-norm} shows that when all the operators in the Hamiltonian are added to the product formula in a single ``add new operators'' procedure, there exists a good solution $\vec{\lambda}'$ such that $\|\vec{\lambda}'\|_1\leq \|a\|_1$ and $\Delta(\vec{\lambda}')\leq \dcut/2$. Therefore, by adding different operators to the product formula, there always exists acceptable $\vlam'$ whose $l_1$ norm is less than or equal to $\|a\|_1$ after $L$ iterations.

Based on the above observations, we give an improved protocol here.

\newcommand{\wcut}{W_{\mathrm{cut}}}
\setcounter{algocf}{2}
\begin{algorithm}[h]
\caption{Adaptive product formula (Regularized)}
\label{alg:joint-regularized}
1. Set $\dcut$, $\wcut$. Input the initial state $\ket{\Psi_0}$ and Hamiltonian $H=\sum_{j=1}^{L} a_j P_j$;

2. (Joint  parameter optimization) Measure $A$ and $C$. Search for a $\vec\lambda$ and that satisfies $\Delta(\vec\lambda)\leq\dcut$ and $\|\vec\lambda\|_1\leq\wcut$ for the current circuit $G(\vec O,\vec\Lambda+\vec\lambda\delta t)\ket{\Psi_0}$ at time $t$. If the search failed, go to step 3. Otherwise, set $\vec\Lambda \rightarrow \vec\Lambda+\vec\lambda\delta t$ and continue step 2 with $t\rightarrow t+\delta t$; Terminate when $t=T$;

3. (Add new operators) 
\begin{enumerate}[(a)]
\item For every Pauli word $P_j$ in $H$ that have not been added in the current ``add new operators'' procedure, do the following procedure on the new circuit $G(\vec O\oplus P_j,(\vec\Lambda+\vec\lambda\delta t)\oplus \lambda'\delta t)\ket{\Psi_0}$.
 \begin{enumerate}
    \item If in the current circuit $\Delta > \dcut/2$: calculate $\vec{\lambda}$ and $\Delta$ of the new circuit without regularization.
    \item If in the current circuit $\Delta \leq \dcut/2$: calculate $\vec{\lambda}$ and $\Delta$ of the new circuit with a regularization such that $\Delta \leq \dcut/2$ is kept and the $\|\vec{\lambda}\|_1$ is minimized.
\end{enumerate}
\item \begin{enumerate}
    \item If in the current circuit $\Delta > \dcut/2$: Set $O^{(\min)}$ to be the Pauli word giving the smallest $\Delta$ in  step (3a).
    \item If in the current circuit $\Delta \leq \dcut/2$: Set $O^{(\min)}$ to be the Pauli word giving the smallest $\|\vec\lambda\|_1$ in  step (3a). 
\end{enumerate}
\item Add $O^{(\min)}$ to the end of the product formula as $\vec{O}\rightarrow \vec{O}\oplus O^{(\min)}$ and $\vec{\Lambda}\rightarrow \vec{\Lambda}\oplus 0$;
\item If we have $\Delta \leq \dcut/2$ and $\|\vec\lambda\|_1<\wcut$, stop adding operators and go to step 2. Otherwise, go to step (3a).
\end{enumerate}
\end{algorithm}

The regularized objective in \autoref{equ:regularized-Delta-obj} can be used in \autoref{alg:joint-regularized} to find $\vlam$ with a small norm.
The value of $\wcut$ should be determined by considering both the shot noise (\autoref{equ:smallest-var-for-delta2}) and the finite time step error (\autoref{equ:delta2}). If $\wcut$ is set to be smaller than $\|a\|_1$, the above protocol might fail because $\|\vec{\lambda}\|_1$ cannot be decreased to $\wcut$. 

One potential problem  in \autoref{alg:joint-regularized} is the measurement cost when $\|\vec{\lambda}\|_1$ has not been reduced. We want to point out that \autoref{thm:lambda-norm-can-decrease} still holds in the presence of noise, as long as the entries of $A$ and $C$ are inherited among iterations (recall that in the proof of \autoref{thm:lambda-norm-can-decrease}, we have $A^{\mathrm{old}}_{ij}=A^{\mathrm{new}}_{ij}$ and $C^{\mathrm{old}}_i=C^{\mathrm{new}}_i$ for $i,j\in\{1,2,\dots,n_O-1\}$). Therefore, though a large $\|\vec{\lambda}\|_1$ induces a large variance, it only makes the operators added to the product formula not optimal and makes the circuit deeper. 

\subsubsection{Scalability analysis}
\label{SM:circuit-scalability}

In this section, we show that the required number of shots is independent of the number of parameters (and operators) in the circuit.  
Let us denote the pessimistic bound of $\var[\Delta'^2(\vlam')]$ in \autoref{equ:var-delta-bound-zero-A-C} by $\var[\Delta'^2(\vlam')]_{\max}$. The bound can be rewritten as
\begin{equation}
\label{equ:max-var-rewritten}
\var[\Delta'^2(\vlam')]_{\max}
= \frac{4}{n_M}\Big(\sum_{i}|\lambda'_{i}|(\sum_{j>i}|\lambda'_{j}|+\|a\|_1)\Big)^2.
\end{equation}
As we pointed in \autoref{SM:reduce-lambda-norm}, the $l_1$ norm of $\vec{\lambda}'$ 
 can be bounded by adding more operators. Next, we study how the variance depends on $\|\vec{\lambda}'\|_1$. Treating  $\|\vec{\lambda}'\|_1$ as an independent constant, we define a Lagrangian function with $K$ being the Lagrange multiplier:
\begin{equation}
\mathcal{L} = \sum_{i}|\lambda'_{i}|(\sum_{j>i}|\lambda'_{j}|+\|a\|_1)- K(\sum_j |\lambda'_{j}|-\|\vec{\lambda}'\|_1).
\end{equation}
which can give $\lambda'_{i}$ for 
$\var[\Delta'^2(\vlam')]_{\max}$
to take its extremum. The derivative of the Lagrangian function is given by
\begin{align}
\frac{\partial \mathcal{L}}{\partial |\lambda'_{k}|} &=  (\sum_{j>k}|\lambda'_{j}|+\|a\|_1) + \frac{\partial \sum_{i<k}|\lambda'_{i}|(\sum_{j> i}|\lambda'_{j}|+\|a\|_1)}{\partial |\lambda'_{k}|} - K\\
&= (\sum_{j>k}|\lambda'_{j}|+\|a\|_1) + \sum_{i<k}|\lambda'_{i}| - K \\
&= \sum_{j\neq k}|\lambda'_{j}|+\|a\|_1-K = \|\vec{\lambda}'\|_1-|\lambda'_{k}|+\|a\|_1-K.
\end{align}
The solution of $\frac{\partial \mathcal{L}}{\partial |\lambda'_{k}|}=0$ gives $|\lambda'_{k}|=\|\vec{\lambda}'\|_1/n_O$, where $n_O$ is the dimension of $\vec{\lambda}'$, i.e., the number of operators in the product formula. Notice that $\frac{\partial^2 \mathcal{L}}{\partial |\lambda'_{k}|\partial |\lambda'_{j}|} \leq 0$ (it can only be $-1$ or $0$), which indicates that we have the maximum here. We have
\begin{align}
&\sum_{i}|\lambda'_{j}|(\sum_{j>i}|\lambda'_{j}|+\|a\|_1)\\
=&\sum_{i}\frac{\|\vec{\lambda}'\|_1}{n_O}(\sum_{j>i}\frac{\|\vec{\lambda}'\|_1}{n_O}+\|a\|_1)\\
=&\frac{n_O(n_O-1)}{2}(\frac{\|\vec{\lambda}'\|_1}{n_O})^2+\|a\|_1n_O\frac{\|\vec{\lambda}'\|_1}{n_O}\\
=&\frac{1}{2}(1-\frac{1}{n_O})\|\vec{\lambda}'\|_1^2+\|a\|_1\|\vec{\lambda}'\|_1
<\frac{1}{2}\|\vec{\lambda}'\|_1^2+\|a\|_1\|\vec{\lambda}'\|_1.
\end{align}
Hence we can have   a bound of the variance as
\begin{align}
\label{equ:smallest-var-for-delta2}
\var[\Delta'^2(\vlam')]_{\max}
< \frac{4}{n_M}(\frac{1}{2}\|\vec{\lambda}'\|_1^2+\|a\|_1\|\vec{\lambda}'\|_1)^2 =\frac{1}{n_M}(\|\vec{\lambda}'\|_1^2+2\|a\|_1\|\vec{\lambda}'\|_1)^2.
\end{align}
Notice that  $\|a\|_1$ does not grow with $n_O$. 
As shown in the last section that there exists $\vec{\lambda}'$ such that $\|\vec{\lambda}'\|_1\leq\|a\|_1$ when enough operators are added to the circuit, the number of shots to achieve a certain accuracy is independent of $n_O$.

\subsubsection{Implication of the result}
\label{SM:failure-probability}

In this section, we show how to estimate the error of the estimation of $\Delta(\vlam')$ and show the measurement cost to achieve a required accuracy of $\Delta'(\vlam')$. For conciseness, we will not specify $\vlam'$ when mentioning $\Delta'^2$, $\Delta^2$, $\Delta'$ and $\Delta$ in this section.

  When there is only statistical noise and $\mathbb{E}[\Delta'^2]=\Delta^2$, by Chebyshev's inequality, we have
\begin{align}
    \Pr(|\Delta'^2-\Delta^2|\geq \sqrt{ \frac{\var[\Delta'^2]}{\delta}}) \leq \delta.
\end{align}
Let us define $\sqrt[*]{~\cdot~}$ such that $\sqrt[*]{|x|}=\sqrt{|x|}$, $\sqrt[*]{-|x|}=-\sqrt{|x|}$. We define the random variable for the estimation of $\Delta$ to be
\begin{equation}
    \Delta' = \sqrt[*]{\Delta'^2},
\end{equation}
so that
\begin{equation}
    \Delta'-\Delta = \sqrt[*]{\Delta^2 + (\Delta'^2-\Delta^2)} -\Delta.
\end{equation}
The above definition of $\Delta'$  makes the analysis easier since $\Delta'^2$ can be negative. By $|\Delta'^2-\Delta^2| < \sqrt{\var[\Delta'^2]/\delta}$, we  have
\begin{align}
    \sqrt[*]{\Delta^2-\sqrt{ {\var[\Delta'^2]}/{\delta}}}  - \Delta  < & ~~ \Delta'-\Delta < \sqrt[*]{\Delta^2+\sqrt{ {\var[\Delta'^2]}/{\delta}}} - \Delta.
\end{align}
To estimate the bound on $\Delta'-\Delta$, we simplify the two sides of the bound by Taylor expansion with the assumption that $\frac{\var[\Delta'^2]}{\Delta^4}$ is small, which yields
\begin{align}
\sqrt[*]{\Delta^2\pm\sqrt{ {\var[\Delta'^2]}/{\delta}}} - \Delta= \Delta\sqrt{1\pm\sqrt{ \frac{\var[\Delta'^2]}{\Delta^4\delta}}} - \Delta = \pm\frac{1}{2}\Delta\sqrt{ \frac{\var[\Delta'^2]}{\Delta^4\delta}} + \Delta \mathcal{O}( \frac{\var[\Delta'^2]}{\Delta^4\delta}).
\end{align}
Therefore, the bound can be simplified as
\begin{align}
    |\Delta'-\Delta| < \epsilon :=\frac{1}{2}\sqrt{ \frac{\var[\Delta'^2]}{\Delta^2\delta}} + \Delta\mathcal{O}( \frac{\var[\Delta'^2]}{\Delta^4\delta}).
\end{align}
This implies that
\begin{equation}
    \Pr\left(|\Delta'-\Delta| \geq \frac{1}{2}\sqrt{ \frac{\var[\Delta'^2]}{\Delta^2\delta}} + \Delta\mathcal{O}( \frac{\var[\Delta'^2]}{\Delta^4\delta})\right) \leq \delta.
\end{equation}
Notice in the preceding sections, $\var[\Delta'^2]$ takes the form of $\var[\Delta'^2] = \frac{V}{n_M}$, where $V$ is a factor related to the Hamiltonian and the current product formula. For example, from \autoref{equ:smallest-var-for-delta2}, $V$ can be $(\|\vec{\lambda}'\|_1^2+2\|a\|_1\|\vec{\lambda}'\|_1)^2$. Thus, we have
\begin{align}
    \epsilon = \frac{1}{2\Delta}\sqrt{ \frac{V}{n_M\delta}} + \Delta\mathcal{O}( \frac{V}{n_M\Delta^4\delta}).
\end{align}
When $n_M$ is large so that the high-order terms are negligible, with the pessimistic assumption that $\|\vec{\lambda}'\|_1=\|a \|_1$ and $V=(\|\vec{\lambda}'\|_1^2+2\|a\|_1\|\vec{\lambda}'\|_1)^2$, we have
\begin{align}
    \epsilon \approx \frac{3\|a\|_1^2}{2\Delta\sqrt{ n_M\delta}}
\end{align}
which grows quadratically with $\|a\|_1^2$. On the other hand, $n_M$ can be approximated by
\begin{align}
    n_M \approx \frac{V}{4\Delta^2\epsilon^2\delta}.
\end{align}
As $V$ can be independent of the number of operators in the Hamiltonian, the measurement cost of our method at each step  can still be small even if there are many operators in the Hamiltonian as long as $\|\vec{\lambda}'\|_1$ and $\|a\|_1$ are small. 
This property implies that our method may still be  useful when simulating the evolution by Hamiltonians, which have many Pauli words yet small $\|a\|_1$. Such Hamiltonians might be found in electronic structure problems. Let $n_{\mathrm{orbital}}$ be the number of orbitals. The number of Pauli words in these Hamiltonians grows as $\mathcal{O}(n_{\mathrm{orbital}}^4)$, but most terms have small strengths and the total weight can be small.
As a cost of shallow circuit depth,  our method is not as scalable as the canonical product formula methods because the required number of measurements scales as $\mathcal{O}(\|a \|_1^4)$ if we assume $\|\vec{\lambda}'\|_1=\|a \|_1$. 
In addition, we note that $\Delta'$ obeys a sub-Gaussian distribution 
and the dependency of $\epsilon$ and $n_M$ on the failure probability $\delta$ could be greatly improved by Hoeffding's inequality.

Finally, let us consider the special case where $\Delta = 0$ as in this case the assumption of $\frac{\var[\Delta'^2]}{\Delta^4}\rightarrow 0$ collapses. 
By the definition $\Delta'=\sqrt[*]{\Delta'^2}$, we have
\begin{equation}
    \Pr(|\Delta'|\geq \sqrt[4]{ \frac{\var[\Delta'^2]}{\delta}}) \leq \delta,
\end{equation}
so that the error on the estimation of $\Delta$ is
\begin{equation}
    \epsilon \approx (\var[\Delta'^2]/\delta)^{\frac{1}{4}} = (V/n_M\delta)^{\frac{1}{4}}.
\end{equation}
The error has a linear scaling on $\|a \|_1$, $\epsilon = \mathcal{O} \left(\|a \|_1{(n_M\delta)^{-\frac{1}{4}} } \right)$, with the assumption of $\|\vec{\lambda}'\|_1\approx\|a\|_1$.

\newcommand{\hadam}{H_{\mathrm{d}}}

\section{Discussions}
\subsection{Comparison to other works}

Compared to the first-order Trotter method, where $O_j$ and $\Lambda_j$ are set to be fixed as the corresponding Pauli word $P_j$ and weight $a_j$ in the Hamiltonian, respectively, our method provides a systematic circuit-growth strategy that optimizes the quantum circuits with much smaller gate counts. 
Our numerical simulation shows a significance improvement of the gate count saving in practical computation.

There are various works that have been proposed to use variational methods to simulate quantum dynamics, which use a  quantum circuit with a size smaller than the circuit of the conventional Trotterization algorithm. 
One major challenge in variational quantum simulation is to design appropriate circuit ansatz to represent the manifold that the target states live in, and in general it is difficult to ensure the simulation accuracy.
Contrary to a pre-determined and fixed circuit ansatz in variational quantum simulation, we optimize the circuits under time evolution by tracking the accuracy of the circuit at each time, which enables an adaptive circuit construction with guarantees on the simulation errors in the time evolution.  
Here, we mainly focus on real-time dynamics, and it would be a future direction to extend the results to imaginary time evolution, which may be used to find the energy spectra of quantum many-body problems, similarly to Refs.~\cite{motta2020determining,sun2021quantum}. Several adaptive variational quantum algorithms have been proposed for finding the energy spectra of quantum many-body problems~\cite{grimsley2019adaptive,tang2021qubit2}, and our results using the quantum Krylov algorithm can be compared.

It is worth comparing our results with a parallel work, termed AVQDS, proposed by Yao \textit{et al} \cite{yao2020adaptive}.
AVQDS is a great work that shares a similar idea to our Protocol 2,  i.e., the jointly-optimized algorithm, which adaptively constructs quantum circuits for real-time evolution.
In contrast, our work is based on the theory of product formula and is proposed with guaranteed simulation accuracy as its first consideration. 
A direct result of this difference is that AVQDS allows more choices of operator pools, whereas our work only allows the selection of operators from the decomposition of the Hamiltonian. This feature of our method   ensures the circuit construction to succeed, which can reduce $\Delta$ to any $\dcut>0$ and achieve a desired simulation accuracy. 
This is important because improper pools may result in algorithmic errors which cannot be further reduced by adding more operators. Also, as we have pointed out in \autoref{SM:reduce-lambda-norm}, by using the Hamiltonian pool the measurement cost can be bounded.

\subsection{Circuit Implementation of $e^{-iPt}$}
In the end, we give a pedagogical explanation of the implementation of $e^{-iPt}$  with a quantum circuit for beginners.   
Assume $P = P_1\otimes P_2 \otimes \cdots \otimes P_n$, where $P_j\in \{X,Y,Z\}$. Notice that $\hadam X\hadam=Z$ ($\hadam$ means the Hadamard gate) and $R_x(\frac{\pi}{2})YR_x(-\frac{\pi}{2})=Z$. We define a map $G$ from a Pauli word to a gate, such that $G(Z)=I,G(X)=\hadam,G(Y)=R_x(-\frac{\pi}{2})$ and $G^{-1}(P_j)P_jG(P_j)=Z$. Then we can decompose $e^{-iPt}$ as 
\begin{equation}
    e^{-iPt} = \exp(-it \bigotimes_{j=1}^n P_j ) = \exp(-it \bigotimes_{j=1}^n G^{-1}(P_j)  Z^{\otimes n} \bigotimes_{j=1}^n G(P_j)) =  \bigotimes_{j=1}^n G^{-1}(P_j) e^{-iZ^{\otimes n}t} \bigotimes_{j=1}^n G(P_j),
\end{equation}
where $G(P_j)$ can be implemented by simple single qubit gates. To implement $e^{-iZ^{\otimes n}t}$, one can first apply a series of CNOT gates
$$
CNOTs = CNOT_{1\rightarrow 2}CNOT_{2\rightarrow 3} \cdots CNOT_{n-1\rightarrow n}
$$
so that, for any input $\ket{\vec z}$ from the computational basis, the last qubit will be transformed into $\ket{z_1\oplus z_2 \oplus \cdots \oplus z_n}$. Then, we can just apply $e^{-itZ}$ on the last qubit and finally apply $CNOTs^{-1}$. An example for the case of $n=4$ is shown below. 
\begin{align*}
\Qcircuit @C=0.8em @R=1.2em {
&\ctrl{1}&\qw   &\qw    & \qw&  \qw&        \qw & \ctrl{1}&\qw \\
&\targ  &\ctrl{1}&\qw   &\qw&   \qw&        \ctrl{1}&\targ&\qw\\
&\qw    &\targ  &\ctrl{1}&\qw&  \ctrl{1}&   \targ&  \qw&\qw\\
&\qw    &\qw    &\targ  &\gate{e^{-it Z}}  &\targ&\qw&\qw&\qw\\
}
\end{align*}
By such construction, one can easily check that, for all inputs $\ket{z}$ from the computational basis, a phase $e^{-it}$ will be applied to it when $z_1\oplus z_2 \oplus \cdots \oplus z_n = 1$ and all the other inputs will remain the same. Thus, the effect of the circuit is equivalent to the operator $e^{-iZ^{\otimes n}t}$. We remark that in practical implementation, the depth of quantum circuits may be further reduced.

As shown above, the operator $e^{-iPt}$ can be realized by applying single-qubit operations to the circuit for $e^{-iZ^{\otimes n}t}$ without introducing additional two-qubit gates. 
In such construction, $2n-2$ CNOT gates will be used.

\end{document}